\documentclass[a4paper,fleqn,usenatbib]{mnras}
\pdfoutput=1 
\usepackage{graphicx,epsfig,amsfonts,amsmath,amssymb,graphicx,morefloats,appendix}
\usepackage{newtxtext,newtxmath}
\usepackage[T1]{fontenc}
\usepackage{ae,aecompl}
\usepackage[figure,figure*]{hypcap}
\usepackage{color}
\usepackage{natbib}
\usepackage[ampersand]{easylist}
\usepackage[caption=false]{subfig}

\setcitestyle{comma}

\newcommand{\degree}{$^{\circ}$}

\title[Detecting H$\alpha$ Emission from the Slug Nebula]{The Detection of Intergalactic H$\alpha$ Emission from the Slug Nebula at $z\sim2.3$}

\author[Leibler et al.]{
Camille N. Leibler,$^{1}$\thanks{E-mail: cleibler@ucsc.edu}
Sebastiano Cantalupo,$^{2}$\thanks{E-mail: cantalupo@phys.ethz.ch}
Bradford P. Holden$^{1}$
and Piero Madau$^{1,3}$
\\
$^{1}$Department of Astronomy and Astrophysics, University of California Santa Cruz, 1156 High Street, Santa Cruz, CA 95064, USA\\
$^{2}$Department of Physics, ETH Zurich, Wolfgang-Pauli-Strasse 27, 8093, Zurich, Switzerland\\
$^{3}$Institut d'Astrophysique de Paris, Sorbonne Universit{\'e}s, UPMC Univ Paris 6 et CNRS, UMR 7095, 98 bis bd Arago, 75014 Paris, France
}

\date{Accepted XXX. Received YYY; in original form ZZZ}

\pubyear{2018}

\begin{document}
\label{firstpage}
\pagerange{\pageref{firstpage}--\pageref{lastpage}}
\maketitle

\begin{abstract}
The Slug Nebula is one of the largest and most luminous Lyman-$\alpha$ (Ly$\alpha$) nebulae discovered to date, extending over 450 kiloparsecs (kpc) around the bright quasar UM287 at z$=$2.283. Characterized by high surface brightnesses over intergalactic scales, its Ly$\alpha$ emission may either trace high-density ionized gas (``clumps") or large column densities of neutral material. To distinguish between these two possibilities, information from a non-resonant line such as H$\alpha$ is crucial. Therefore, we analyzed a deep MOSFIRE observation of one of the brightest Ly$\alpha$ emitting regions in the Slug Nebula with the goal of detecting associated H$\alpha$ emission. We also obtained a deep, moderate resolution Ly$\alpha$ spectrum of the nearby brightest region of the Slug. We detected an H$\alpha$ flux of F$_{\rm{H\alpha}}= 2.62\pm 0.47 \times 10^{-17}$ erg/cm$^2$/s (SB$_{\rm{H\alpha}}=2.70\pm 0.48 \times 10^{-18}$ erg/cm$^2$/s/\sq\arcsec) at the expected spatial and spectral location. Combining the H$\alpha$ detection with its corresponding Ly$\alpha$ flux (determined from the narrow-band imaging) we calculate a flux ratio of $\rm{F_{Ly\alpha}}/\rm{F_{H\alpha}}=5.5\pm1.1$.  The presence of a skyline at the location of the H$\alpha$ emission decreases the signal to noise ratio of the detection and our ability to put stringent constraints on the H$\alpha$ kinematics. Our measurements argue for the origin of the Ly$\alpha$ emission being recombination radiation, suggesting the presence of high-density ionized gas. Finally, our high-resolution spectroscopic study of the Ly$\alpha$ emission does not show evidence of a rotating disk pattern and suggest a more complex origin for at least some parts of the Slug Nebula.

\end{abstract}


\section{Introduction}
\label{Intro}

In the standard paradigm of galaxy formation and evolution, galaxies are thought to be fueled by accreting material from their surrounding circumgalactic medium (CGM). However, the properties of this accreting material, such as the density, temperature, angular momentum and morphology, remain uncertain. Some cosmological simulations suggest that most of this material accretes in the form of relatively cold (T$\sim10^{4}$K) intergalactic filaments. This has even been found to be the case for the most massive galaxies at high redshift, for which a stable hot corona should be in place \citep{Dekel2009}. On the other hand, theoretical arguments and higher resolution simulations have highlighted that such streams may not be able to survive instabilities \citep{Nelson2013,Mandelker2016}. Alternatively, such material could result from the cooling of the hot corona \citep{Voit2015}. In order to distinguish between these two scenarios, direct imaging of the CGM and intergalactic gas is essential. 

Unfortunately, the expected emission of both the cold component (due to the recombination radiation of gas ionized by the cosmic ultraviolet background) 
and hot component (due to X-ray bremsstrahlung) of the CGM around a typical galaxy at z>2 is well below current detection limits \citep[e.g.][]{Cantalupo2005,Gallego2017}. Local ultraviolet (UV) radiation fields, such as in the vicinity of a bright active galactic nucleus (AGN), may be used to increase the detectability of Ly$\alpha$ emission. Indeed, in recent years, several enormous Ly$\alpha$ nebulae (ELANe) have been discovered at z >2 around bright radio-quiet quasars. ELANe are characterized by their extended Ly$\alpha$ emission that traces the CGM, and even intergalactic medium (IGM), out to several hundred kpc from their quasars. These detections were made using custom-made narrow-band (NB) filters on the W.M. Keck telescope \citep{Cantalupo2014, Hennawi2015} or by performing integral field spectroscopy using MUSE on the ESO/VLT (\citealt{Borisova2016, FAB2018}; see \citealt{Cantalupo2017} for a review).

The largest and brightest of such Ly$\alpha$ emitting structures, nicknamed the ``Slug Nebula", was discovered by \citet{Cantalupo2014}. The ``Slug" was found near the radio-quiet quasar UM287 using a custom NB filter on the Low Resolution Imaging Spectrometer (LRIS) instrument mounted on the Keck I telescope. With a total projected size of at least 480 physical kpc, this nebula extends well beyond the virial radius of the halo of a typical bright quasar host with a mass of $\sim10^{12.5}\rm{M_\odot}$ (see \citealt{daAngela2008} and \citealt{Trainor2012}). The Slug Nebula, therefore, represents the best system available to date in which to jointly study the circumgalactic and intergalactic medium in emission. 

The filamentary and asymmetric morphology of the Slug Nebula is similar to the predictions of recent cosmological simulations. However, the very high surface brightness (SB) of the Ly$\alpha$ emission (above 10$^{-17}~\rm{erg~s}^{-1} ~\rm{cm}^{-2} ~\rm{arcsec}^{-2}$) extending over hundreds of kpc presents a serious challenge to our current theoretical understanding of baryonic structure formation in the massive halos associated with quasars.  As discussed in \citet{Cantalupo2014}, there are at least two possible scenarios for the origin of the extended Ly$\alpha$ emission: i) fluorescent Ly$\alpha$ emission following hydrogen recombinations of the gas ionized by the quasar, and ii) Ly$\alpha$ ``photon-pumping" or ``scattering" of the quasar broad line region emission.

In the first case, the observed Ly$\alpha$ SB can only be explained if the recombining gas is ``cold" (T<10$^{5}$ K) and has very large densities ($>1-10$ cm$^{-3}$) that are much higher than the typical gas densities expected at such large distances from a galaxy. However, because recombination emission scales with the density squared, a small volume filling factor or a large gas clumping factor (C>1000) below the scale of a few kpc could explain the Ly$\alpha$ emission as well as the much lower volume-averaged densities. Therefore, this interpretation of the data would require dense photoionized ``clumps" of gas within the CGM but these ``clumps" must have sizes that are well below the current resolution limits of cosmological simulations (see e.g., \citealt{Cantalupo2014}, \citealt{FAB2015} for further discussion).

The second case, as discussed in \citet{Cantalupo2014}, would require very large column densities of neutral gas above $10^{20}$ cm$^{-2}$ (i.e., corresponding to damped Ly$\alpha$ systems or DLAs) to be present on scales of hundreds of kpc around the quasar. This material would then have to be illuminated by the Ly$\alpha$ emission of the quasar's broad line region without being photoionized by the quasar itself. Although optically thick gas is routinely observed in the proximity of quasars \citep{Prochaska2013}, such large column densities of neutral material over these distance scales are not typically observed either in absorption studies or in cosmological simulations. However, recent deep observations have found that some DLAs are also associated with Ly$\alpha$ nebulae, although these nebulae have much smaller physical scales and luminosities than the Slug \citep{Fumagalli2017}. 

Either scenario therefore requires very high densities and Ly$\alpha$ observations alone are not able to distinguish whether the gas is mostly neutral and diffuse or ionized and clumpy.  In order to break this degeneracy, a non-resonant line such as \ion{He}{II}[1640] or H$\alpha$ is needed. In particular, a detection or limit on H$\alpha$ emission would put the most stringent constraint on the recombination or ``scattering" origin of the Ly$\alpha$ emission because these two transitions arise in the same atom. Another advantage of searching for H$\alpha$ emission is that that the presence of \ion{He}{II}[1640] emission could require favourable conditions in terms of the ionization spectrum and ionization parameter (see e.g., \citealt{FAB2015}; Cantalupo et al., in prep.). 

To our knowledge, there have been no reported detections prior to our study of H$\alpha$ emission from intergalactic gas associated with Ly$\alpha$ nebulae around radio-quiet quasars.\begin{footnote}{Detection of H$\alpha$ emission from galaxies embedded in some Ly$\alpha$ nebulae at z$\sim$2.3 have been reported by, for example \citet{Yang2014}}\end{footnote} Similarly, no \ion{He}{II}[1640] emission has been found in long-slit spectroscopic observations of ELANe around radio-quiet quasars \citep{FAB2015}, although deep integral-field spectroscopic observations are now revealing HeII emission at fainter levels than expected (Cantalupo et al., in prep.). In contrast with ELANs, a few detections of \ion{He}{II}[1640] in radio-quiet Ly$\alpha$ blobs (LABs, see \citealt{Cantalupo2017} for a review) have been reported (e.g., \citealt{Prescott2015}), though the majority of LABs show no sign of \ion{He}{II}[1640] emission \citep{FAB2015a}. 

Though the terms LAB and ELAN are often used interchangeably, it is important to note some distinctions. ELANe are bright ($\rm{L_{Ly\alpha}}\sim10^{44}$ erg/s) Ly$\alpha$ nebulae around z>2 quasars with extents >100 kpc (e.g. \citealt{Cantalupo2014,Hennawi2015,FAB2018}). Though comparable in size and brightness to ELANe, LABs were historically distinguished by their apparent lack of association with an AGN or bright continuum source at the time of their discovery (e.g., \citealt{Steidel2000, Matsuda2004, Dey2005, Prescott2009, Yang2009, FAB2015a}). However, follow-up observations of LABs often uncovered evidence of the presence of obscured AGN or massively star-forming galaxies (e.g., \citealt{Chapman2001,Geach2009,Overzier2013,Prescott2015b,Hine2016}). Therefore, the term LABs has started being used by some authors to refer to large Ly$\alpha$ nebulae with physical extents greater than $\sim 100$ kpc.

Given this broader definition of LABs, ELANe could be considered a subtype of LABs. However, since LABs encompass a wide variety of systems, it would be a mistake to blindly apply any inferences about ELAN emission mechanisms to LABs as a whole. Similarly, though very extended Ly$\alpha$ nebulae have been found around high redshift radio-loud galaxies with large radio jets, these more commonly exhibit extended \ion{He}{II}[1640] emission and broader kinematics. This could suggest that different processes are at play between radio-quiet and radio-loud systems (see, e.g., \citealt{VillarMartin2007}, \citealt{Miley2008}, and \citealt{Cantalupo2017} for reviews). 

Nevertheless, despite the different classifications and nomenclatures associated with highly extended Ly$\alpha$ emission discovered in the last decades, they are almost always associated with AGN or massively star forming galaxies. This suggests that the presence of a strong ionizing field, and therefore emission produced by fluorescent recombination radiation, is likely a necessary requirement in all cases (see \citealt{Cantalupo2017} for discussion).

In this paper, we report the results of our search for extended H$\alpha$ emission from the Slug Nebula using long-slit near-IR spectroscopy with the new Multi-Object Spectrometer For Infra-Red Exploration (MOSFIRE) instrument on the W.M. Keck I telescope. We also perform high-resolution Ly$\alpha$ spectroscopy of a similar region in the Slug Nebula with the goal of both guiding our H$\alpha$ search in the velocity dimension and gaining a deeper understanding of the Ly$\alpha$ kinematics. 

The paper is organized as follows:
In \S\ref{sec:LRISspec}, we describe the deep Keck I/LRIS observations taken of the brightest region of the Slug Nebula and in \S \ref{sec:LRISredux}, discuss the data reduction process of these observations. Similarly, the two nights of deep near-infrared spectroscopy, obtained using Keck I/MOSFIRE , and their reduction, are described in \S 2.3 and \S 2.4 respectively. In \S \ref{sec:LRISkinematics} we explore the Ly$\alpha$ kinematics of the Slug and in \S \ref{sec:Lyaflux} we measure the Ly$\alpha$ flux contained within the MOSFIRE N1 slit. \S \ref{sec:Haemission} and \S \ref{sec:Halphaflux} discuss the measurement of the Slug Nebula's H$\alpha$ flux. We also extract the 1-D spectra of two compact sources in the vicinity of the Slug Nebula (z$\sim$ 2.287) from the MOSFIRE N1 and N2 observationsn in \S \ref{sec:compactsources}, and calculate their \ion{N}{II} and H$\alpha$ fluxes.  In \S \ref{sec:DiscussLyaKinematics} we examine the implications of the Ly$\alpha$ kinematics for the Nebula's gas distribution. In \S \ref{sec:emissionmech}, we compute the Ly$\alpha$ to H$\alpha$ flux ratio and compare it to predictions from case B recombination radiation. \S In \S \ref{sec:DiscussCompact}, we constrain the origin of the compact source ``C" and ``D" emission (AGN vs star-formation vs QSO A fluorescence). Finally, in \S \ref{sec:Conclusions}, we summarize our results.

\ \\
\section{Observations}
\label{Data}

\subsection{LRIS Spectroscopy}
\label{sec:LRISspec}

On UT September 09, 2015, we used the blue camera of the Low Resolution Imaging Spectrometer \citep[LRIS;][]{LRIS} on the Keck I 10m telescope to observe the Ly$\alpha$ emission of the UM287 nebula (also referred to as the ``Slug Nebula" or ``Slug"). The spectra were obtained with a 1" slit as part of a multi-object slit mask. The slit was oriented with a position angle (PA) of 322\degree, to match the PA of the MOSFIRE Night 2 (N2) mask (see Figure \ref{fig:SlitPA}). In order to cover the Ly$\alpha$ emission of the Slug, we used the D460 dichroic and the 1200 lines mm$^{-1}$ grism blazed at 3400{\AA}, which covers $\approx$3300$-$4200{\AA}. The measured full-width at half-max (FWHM) was found to be $\sim$ 1". We acquired 9$\times$1800s science exposures, for a total exposure time of 4.5 hrs. In between each exposure, we dithered $\sim$ 1" along the slit. In addition to the science exposures, we took bias frames, arcs, as well as slitless and slitted twilight flats which were used in the data reduction process. All exposures were readout with 1x1 CCD binning.

\subsection{LRIS Calibrations and Data Reduction}
\label{sec:LRISredux}

The LRIS blue camera data were reduced using the publicly available LowRedux package, distributed within XIDL \citep{XIDL} 
producing nine calibrated, unfluxed 2D spectra. This pipeline performs standard data reduction steps, including overscan and bias subtraction, flat fielding, and wavelength calibration.

The flat fielding procedure constructs a pixel flat used to correct for pixel sensitivity variation from the slitless twilight flats. In addition, the slitted twilight flats are utilized to correct for the non-uniform illumination of the slit. LowRedux determines a wavelength solution by fitting low order polynomials to the arc lamp spectra and is reported in air wavelengths for the 2D spectra. 

We wrote custom python scripts using the Astropy \citep{Astropy}, IPython \citep{IPython}, Matplotlib \citep{Matplotlib}, NumPy \citep{NumPy}, and SciPy \citep{SciPy} packages, to coadd the individual reduced spectra since LowRedux does not combine 2D spectra. Due to the dithering along the slit between exposures, each image needed to be shifted, in the spatial direction, to a common frame (chosen to be that of the fifth exposure). To calculate the required shift, we fit gaussians to the spatial profile of a star in a separate slit on the mask, and determined the change in the centroid position. The applied shifts were rounded to the nearest integer pixel in order to avoid interpolation and complications in calculating the associated error. The uncertainty associated with integer shifts is at most 1/2 pixel, which corresponds to an error of 0.0675\arcsec, which is far less than the 1\arcsec\ seeing disk.

LRIS is known to experience significant telescope pointing position dependent flexure, which shifts the location of a fixed wavelenth on the detector. LowRedux uses the known wavelengths of skylines to measure, in an extracted 1D spectrum, the wavelength solution offset caused
by the flexure. We therefore used LowRedux to extract a 1D spectrum of compact source C (see Figure \ref{fig:LRIS2Dspec}), which is located in the same slit as the Slug Nebula, and calculate the flexure-induced spectral pixel shift for each exposure. This shift of $\sim$11-13 unbinned pixels was then rounded to the nearest integer pixel and applied to the spectral direction of each of the 2D spectra. The rounding error of at most 1/2 pixels amounts to an uncertainty of $\approx0.135$\text{\AA} or $\approx$10 km/s. 

Once the exposures were corrected for flexure and dithering offsets, we ran each of them through the publicly distributed ``dcr" package \citep{DCRpaper,DCRcode} to detect and remove cosmic rays. We then coadded these nine cleaned images by summing the electron counts in each pixel and renormalizing by the total exposure time. In addition, the corresponding 2D wavelength fits file produced by LowRedux was converted from air wavelengths to vacuum wavelengths. 

Lastly, we flux calibrated our coadded spectrum using the deep narrow band (NB) imaging of the UM287 field presented in \citet{Cantalupo2014}.  
In order to do so, we applied the LRIS slit to the NB image, choosing only the pixels that contributed to the flux in our spectrum. Next, we trimmed off the outer edges of the slit, selecting only the region of the spectrum that has flux from the nebula and very good background subtraction. We then summed up the NB flux within this shortened slit, which was 4 binned pixels wide by 135 binned pixels long and covered 1.08" by 36.45" on the sky.

This total NB flux is compared to that of the 2D spectrum over the same wavelength range and spatial location. The equivalent flux of the 2D spectrum is calculated by first applying the filter transmission function to the spectral direction of the spectrum. We then integrate the flux, in e$\rm{^{-}}$/s, over the shortened slit region and divide the total NB flux by the summed 2D spectrum flux to compute the conversion factor from e$\rm{^{-}}$/s to erg/cm$^2$/s. This conversion factor is then applied to each pixel of the LRIS spectrum to produce a fully flux-calibrated 2D spectrum. 

\begin{figure*}
\centering
\hspace*{\fill}%
\subfloat[]{\includegraphics[width=0.5\textwidth]{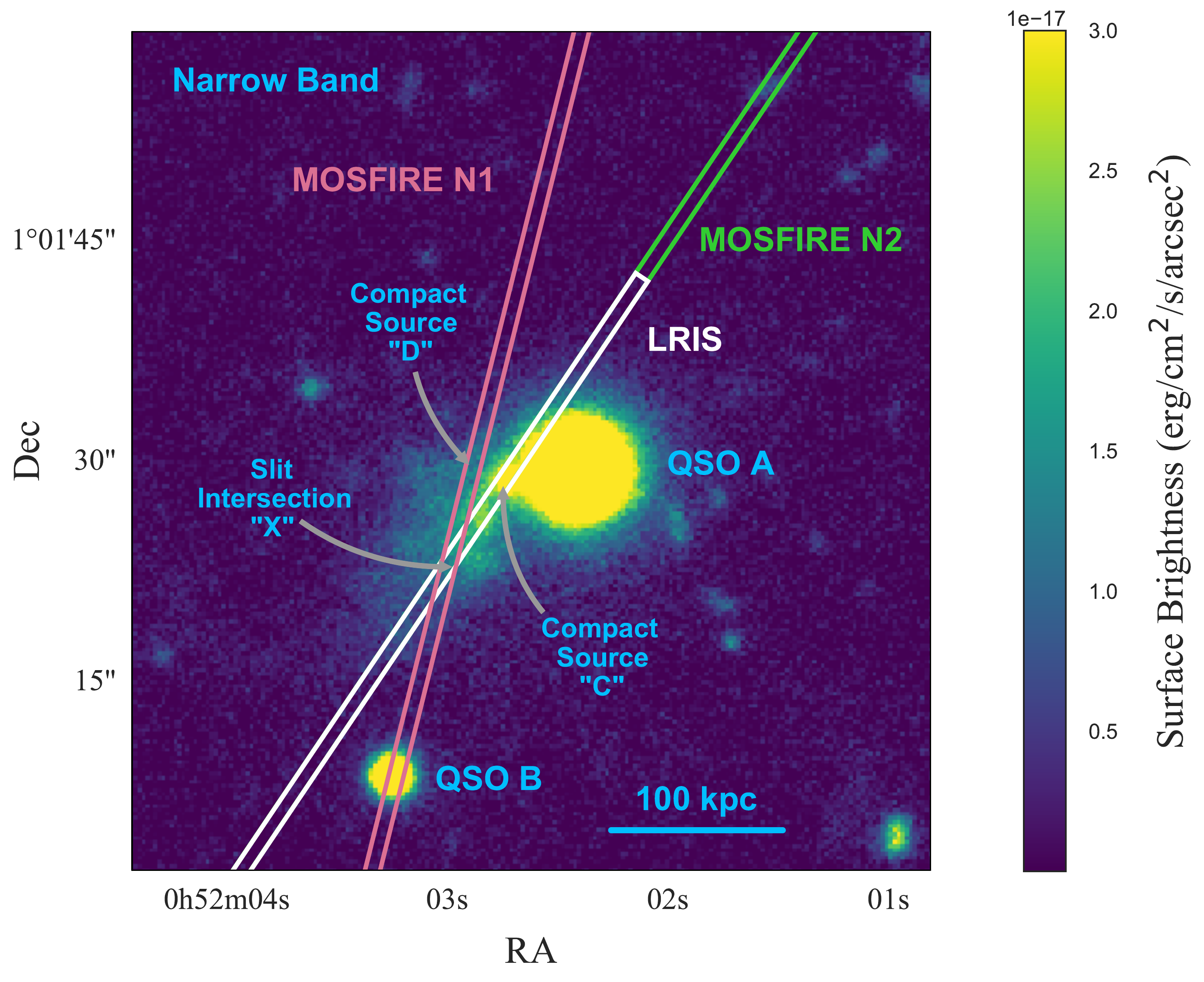}}
\hfill
\subfloat[]{\includegraphics[width=0.5\textwidth]{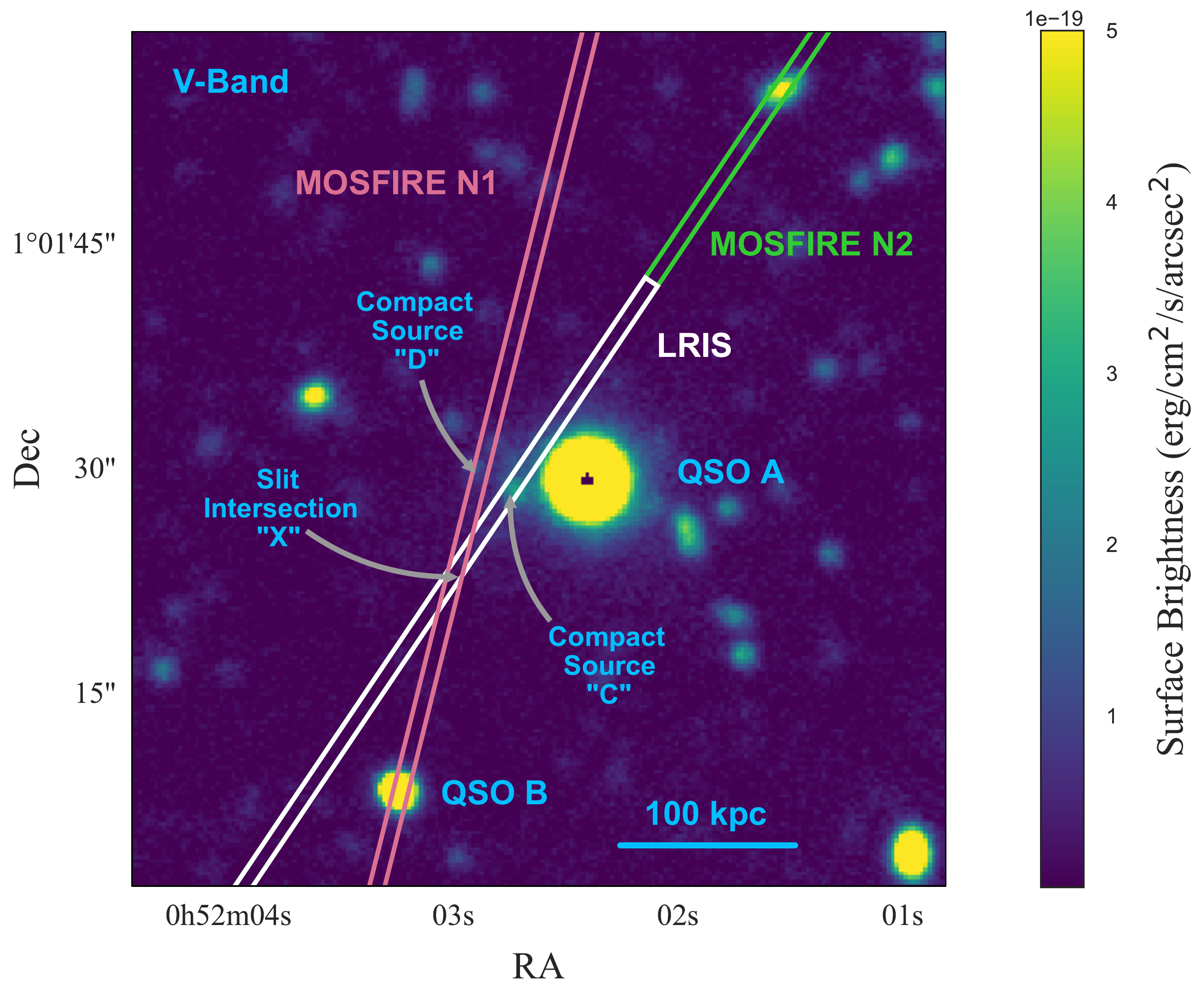}}
\hfill

\caption{This figure depicts the 10 hr narrow-band image (left panel) and deep continuum image (right panel) of the region surrounding the UM287 Nebula, adapted from Figure 1 of \citet{Cantalupo2014}. The narrow-band data was taken on the Keck I telescope with the LRIS NB3985 filter, selected to cover Ly$\alpha$ emission at the redshift of QSO A. The continuum image was taken simultaneously with the LRIS V-band filter on the red camera. The MOSFIRE Night 1 (N1), MOSFIRE Night 2 (N2) and LRIS slit positions are overlaid in red, green, and white respectively. The green MOSFIRE N2 slit extends the full length of the image, overlapping with the white LRIS slit.  QSOA, QSOB, the compact source ``C", the compact source ``D", and position ``X" (the intersection of the MOSFIRE N1 and LRIS slits), are also labeled.}
\label{fig:SlitPA}
\end{figure*}

In order to use an integer number of pixels, the width of the NB shortened slit corresponds to 1.08" which is slightly bigger than the LRIS slit width of 1". Therefore, we expect the flux-calibration of the LRIS spectrum to be biased slightly high, by approximately 8\%. To estimate the systematic error on our flux calibration, we calculated the flux of the compact source, marked as ``C" in Figures \ref{fig:SlitPA} and \ref{fig:LRIS2Dspec}. We find that the compact source flux in the NB and spectrum differ by 20$\%$, which we will take to be our systematic uncertainty.

\begin{table}
\caption{Coordinates for the MOSFIRE Mask Targets. We also include here the coordinates of QSO A for completeness.}
\label{tab:MOScoords}
 \begin{tabular}{lllc}
  \hline
Target Name & RA (J2000) & Dec (J2000) & Slitlet $\#$  \\ 

  \hline
 \vspace{+0.2cm} 
2MASS Star 1 & $00^{\rm{h}}51^{\rm{m}}54^{\rm{s}}.26 $ & $+01\degr 03\arcmin21\farcs2$ & 1 \\ \vspace{+0.2cm} 
QSO A & $00^{\rm{h}}52^{\rm{m}}02^{\rm{s}}.40 $ & $+01\degr 01\arcmin29\farcs3$  & N/A \\ \vspace{+0.2cm}
UM287 Nebula & $00^{\rm{h}}52^{\rm{m}}02^{\rm{s}}.99 $ & $+01\degr 01\arcmin23\farcs1$  & 2 \\ \vspace{+0.2cm}  
QSO B (Night 1 only) & $00^{\rm{h}}52^{\rm{m}}03^{\rm{s}}.26 $ & $+01\degr 01\arcmin08\farcs6$  & 2 \\ \vspace{+0.0cm}  
2MASS Star 2 & $00^{\rm{h}}52^{\rm{m}}07^{\rm{s}}.78 $ & $+00\degr 59\arcmin06\farcs9$& 3 \\ 
   \hline
  \end{tabular}
\end{table}

We observed the Slug Nebula on October 02, 2014 (Night 1 or N1) and October 03, 2014 (Night 2 or N2), using the Multi-Object Spectrometer for Infra-Red Exploration \citep[MOSFIRE;][]{McLean2010,McLean2012} on the Keck I 10m telescope. The spectra were taken using the K-band grating so as to cover the expected H$\alpha$ emission ($\lambda6562.8$) of the nebula (z$\sim$2.283), with a total wavelength coverage of 19540-24060{\AA}. We used a 1" slit width for both nights of observation, resulting in a spectral resolution of R $\sim$ 2500 at $\lambda =$21545.67{\AA}. 

The two masks we designed had three slitlets; the middle slitlet was centered on a region of the UM287 nebula predicted to have the highest H$\alpha$ emission, while the top and bottom slitlets were aligned on two 2MASS stars (all coordinates are shown in Table \ref{tab:MOScoords}). These 2MASS stars were included to help locate the exact position of the UM 287 nebula in case of a non-detection as well as help track the drift of the mask across the detector \citep[see][]{Kriek2015}. 

For the first night, we used a slit at a position angle (PA) of 342\degree. We observed using an AB'BA' dither pattern with offsets of $+51", -17", +17", -51"$ respectively and exposure times of 119.3s for a total integration of 4.8hrs. The median seeing during Night 1 was about 0.7" for both nights but there were intermittent cirrus clouds such that the conditions were not photometric. So as to cover a larger area of the nebula, we used a different PA for Night 2 of 322\degree, centered on the same patch of nebula as for Night 1 (see Figure \ref{fig:SlitPA}). We observed using the same AB'BA' dither pattern and exposure times as for Night 1, for a total integration of 2.6 hrs. Both slit orientations are shown in Figure \ref{fig:SlitPA}.

\subsection{MOSFIRE Calibrations and Data Reduction}
\label{sec:MOSFIREredux}

At the beginning of each night, we took neon and argon arcs through the masks as well as dome flats and thermal flats. Immediately prior to observing UM287, we took spectroscopic standards of HIP5164 with a 1" longslit on Night 1 and a 0.7" longsit on Night 2. These were used to flux calibrate our data as well as correct for telluric absorption. Since we only observed HIP5164 once in the evening, we cannot account for any changes to the telluric absorption throughout the night.

The UM287 data as well as the standard HIP5164 were reduced using the publicly available MOSFIRE Data Reduction Pipeline (version August 2016) \citep[DRP;][]{Steidel2014}. The DRP first flat fields the images and traces the slit edges. To correct for the dome's emission of K-band wavelength photons, the software subtracts the thermal flats from the dome flats before creating a normalized combined flat.  Next, the code combines exposures and preforms the wavelength calibration, combining an interactive fitting of the night sky lines with neon and argon arcs to correct for the faintness of the sky lines at the reddest wavelengths. The sky background is then subtracted and the images are rectified, producing a 2D spectrum for each slitlet along with their corresponding noise frames and integration time maps. 

To prevent smearing out of the emission due to the mask drift across the detector over the course of our observations (see for instance, \citealt{Kriek2015}), we reduced our Night 1 data in 6 batches of 24 exposures ($\sim$ 48 mins) each (other than the last batch which only had 23 exposures).  We then measured the mask drift between batches by tracking the centroids of the two 2MASS stars (see Table \ref{tab:MOScoords}) also present in the MOSFIRE mask. We found shifts of $\pm$1-2 pixels (0.18$\arcsec$-0.36$\arcsec$) between batches which we corrected for before coadding the data. Due to the shortness of the Night 2 observations and the shifts of only about 1 pixel found in the Night 1 data over the course of $\approx$ 1hr, we did not bother correcting for mask drift in the Night 2 data. 

In order to flux calibrate our UM287 nebula 2D spectrum, we used the spectrum of the A0 standard star HIP5164. The 1D spectrum of HIP5164 was derived from the 2D spectrum returned by the DRP, using a boxcar extraction that assumes a gaussian spatial profile. We then calculated the sensitivity function by comparing this 1D spectrum to a template spectrum of Vega from \citet{Bohlin2014} that has the near-infrared emission from the debris disk removed. That spectrum was then renormalized to have the same 2MASS magnitude as HIP5164. This sensitivity function as well as a simple slit loss correction to account for the finite slit width were applied to achieve the final flux calibrated UM287 nebula 2D spectrum, shown in Figure \ref{fig:LRIS2Dspec}. 

\section {Analysis and Results}

\begin{figure}
\begin{center}
\includegraphics[width=0.5\textwidth]{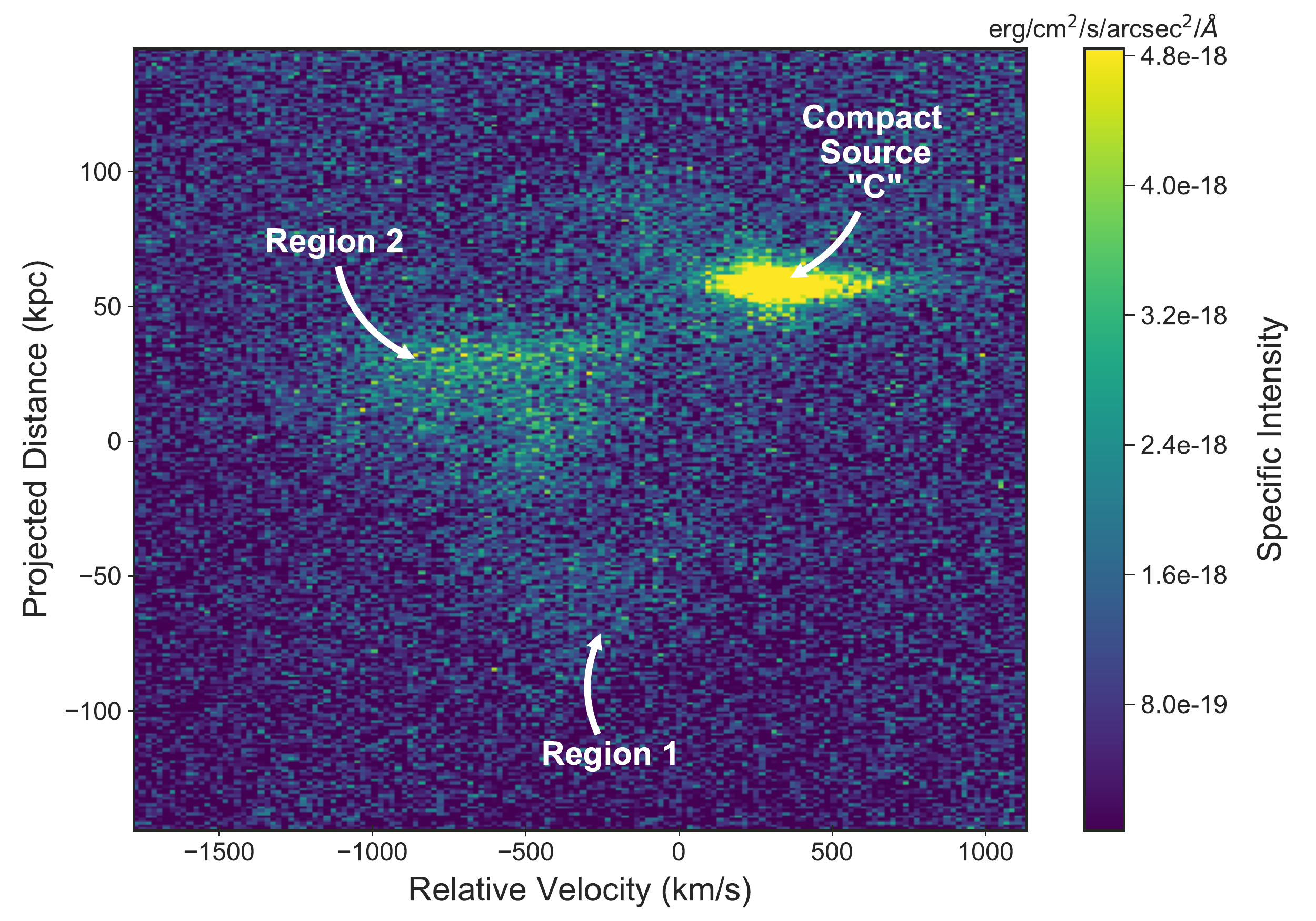}

\end{center}
\caption{The unsmoothed two-dimensional spectrum taken with Keck I LRIS (the white slit in Figure \ref{fig:SlitPA}). The $v=0$ km/s corresponds to the expected Ly$\alpha$ emission at a redshift of z=2.283, the redshift of QSO A. The projected distance corresponding to 0 kpc indicates the location of the MOSFIRE N1 and LRIS slit intersection, referred to as position ``X" in  Figure \ref{fig:SlitPA}. The bright Ly$\alpha$ emitter around $v\sim 300$ km/s and spatial position of $\sim 60$ kpc is the compact source ``C" (see Figure \ref{fig:SlitPA}). The Slug Nebula has a physical extent along the slit of $\sim 150$ kpc.  Its Ly$\alpha$ emission is blue-shifted with respect to that of the compact source (z=2.287) and the redshift of QSO A (z=2.283). }
\label{fig:LRIS2Dspec}
\end{figure}

The final 2D LRIS and MOSFIRE spectra are shown in Figures \ref{fig:LRIS2Dspec} and \ref{fig:Mosfire2Dspec}. In this section, we first examine the kinematics of the Slug Nebula. We use the LRIS spectrum from \S \ref{sec:LRISspec}  which will provide the kinematic structure, which we can use to measure the distribution of the gas. We will use simple moments to characterize the distribution as a way of quantifying how the material is moving with respect to the QSO we assume is illuminating the Slug. Second, we will measure the amount of flux coming from the Slug in the MOSFIRE spectra. In particular, we will measure the flux in H$\alpha$. This result will help us constrain the mechanism that produces the emission. 

\subsection{The Ly$\alpha$ kinematics}
\label{sec:LRISkinematics}

\begin{figure*}
\begin{center}
\includegraphics[width=\textwidth]{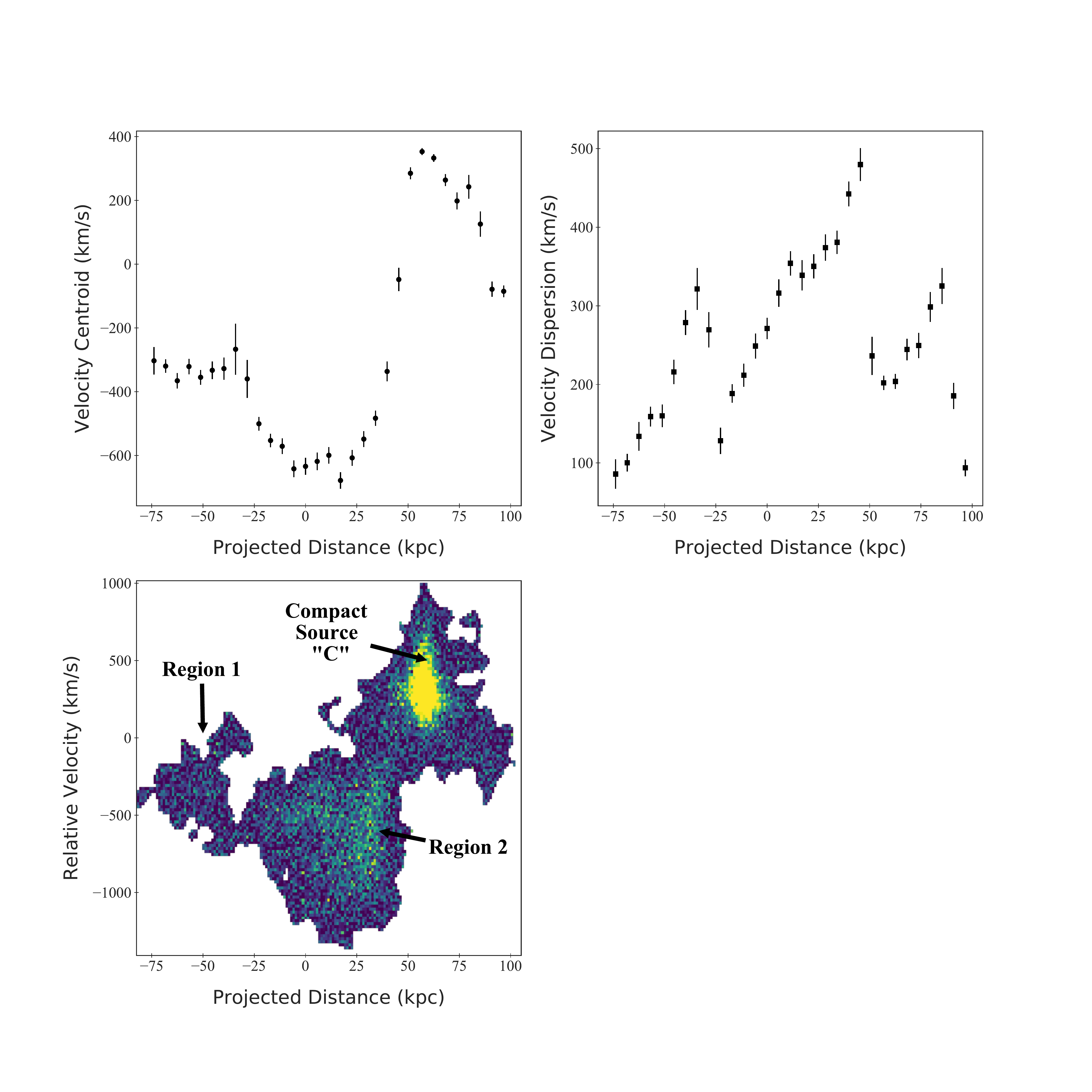}
\end{center}

\caption{The Ly$\alpha$ kinematics of the region around the Slug Nebula. The flux-weighted first moment of the velocity distribution, also referred to as the velocity centroid, is plotted as a function of position along the slit in the top-left panel. The velocity centroid was calculated according to Equation \ref{eq:centvel}, within spatial bins of 5 pixels. Similarly, the flux-weighted second moment of the velocity distribution about the-flux weighted mean, which we refer to as the velocity dispersion, was calculated according to Equation \ref{eq:veldisp}. The velocity dispersion in spatial bins of 5 pixels is shown as a function of projected distance along the slit in the top-right panel of this figure. In both top panels, the 1$\sigma$ error bars were computed using standard bootstrapping techniques. The area of Ly$\alpha$ emission used to measure the velocity centroid and dispersion only includes pixels with a SNR $\geq3$, and is depicted in the bottom-left panel (the color bar matches that of Figure \ref{fig:LRIS2Dspec}).  The $v=0$ km/s corresponds to the expected Ly$\alpha$ emission at the redshift of QSO A (z=2.283). The projected distance of 0 kpc indicates the location of the MOSFIRE N1 and LRIS slit intersection.Three distinct spatial regions, with different kinematic properties, are apparent in both the velocity centroid and velocity dispersion plots: the dimmer left-most region of the Nebula (``region 1"), the brighter region of the Nebula to the right (``region 2") and the compact source ``C'' area in the right-most part of the figure.  
}
\label{fig:LyaKinematicsALL}
\end{figure*}

The velocity centroid and velocity dispersion as a function of spatial position for the Ly$\alpha$ emission around the Slug Nebula is shown in the top left and top right panels of Figure \ref{fig:LyaKinematicsALL}, respectively. To calculate these kinematic tracers, we first selected an appropriate region around the Slug Nebula. We started by running CubExtractor (Cantalupo, in prep.; see also \citealt{Borisova2016} and \citealt{Marino2017} for a short description) on the LRIS 2D spectrum with spatial and wavelength Gaussian smoothing of ($\sigma=2$ pixels) and a signal to noise threshold of 3 per smoothed pixel. The resulting region is shown in the bottom left panel of Figure \ref{fig:LyaKinematicsALL}. 

Next, we partitioned the spatial extent of Ly$\alpha$ emission into bins of 5 pixels (5.68 kpc). Note that these are not independent regions since the seeing in the spatial direction was $\sim 10$ pixels but this box size allows us to finely sample the kinematics of the transition region between the Slug Nebula and the compact source. We then calculated the flux-weighted first and second moments of the Ly$\alpha$ velocity distribution for each spatial bin according to equations \ref{eq:centvel} and \ref{eq:veldisp}. 
\begin{equation}
\label{eq:centvel}
V_{\rm{cent}} = \frac{\sum{vF(v)}}{\sum{F(v)}}
\end{equation}

\begin{equation}
\label{eq:veldisp}
V_{\rm{disp}} = \sqrt{\frac{\sum{(v-V_{\rm{cent}})^{2}F(v)}}{\sum{F(v)}}}
\end{equation}

The flux-weighted mean velocity, which we also refer to as the centroid velocity,  is presented in the left panel of Figure \ref{fig:LyaKinematicsALL} as a function of spatial distance from point X (the intersection point of the Night1 and Night 2 slits denoted in Figure \ref{fig:SlitPA}). The standard error on the velocity centroid, for each spatial bin, was determined using statistical bootstrapping and is shown as the black error bars. Since each spatial bin is about half the size of the atmospheric seeing, each of these bins are correlated. Therefore, the bootstrapped errors are likely an underestimate of the true errors. 

As seen in the top left panel of Figure \ref{fig:LyaKinematicsALL}, the kinematics indicate that the Slug Nebula, the spectrum between 35 kpc and $-$75 kpc, is comprised of two regions with distinct kinematics. ``Region 1'' is at $\sim-$50 kpc has a velocity centroid of $-333\pm12$ while the ``region 2'' is at $\sim$25 kpc is centered at $-555\pm$8. There is then a sharp transition around $\sim$40 kpc marking the beginning of the compact source, which is centered at a velocity of 254$\pm$8 km/s. 

The top right panel of Figure \ref{fig:LyaKinematicsALL} shows the flux-weighted velocity dispersion of the Ly$\alpha$ emission as a function of the projected distance from point X. The corresponding error bars for the velocity dispersion were computed using statistical bootstrapping. As with the velocity centroid, the velocity dispersion of the Slug Nebula displays the same demarkations between the two regions that comprise the Slug and the compact source ``C". Their representative velocity dispersions are 217$\pm$7 km/s, 418$\pm$6 km/s, 453$\pm$9 km/s respectively.

\subsection{The Ly$\alpha$ Flux of the Slug Nebula}
\label{sec:Lyaflux}

We used the narrow band image (see the top panel of Figure \ref{fig:SlitPA}) to calculate the Ly$\alpha$ flux of the Slug Nebula in the region defined within the MOSFIRE Night 1 slit, corresponding to where we measure the H$\alpha$ emission in \S \ref{sec:Halphaflux}. So that comparisons to the MOSFIRE data would be as accurate as possible, we chose the same spatial width and centroid as was used to compute the H$\alpha$ flux (see \S \ref{sec:Halphaflux}). Though we could not precisely select the same velocity range, comparing the narrow band and continuum images shows that there are no continuum sources that could be contaminating the Ly$\alpha$ measurement in our region of interest. In addition, the narrow band filter covers a much larger spectral window than the velocity dispersion of the nebula (see Figure \ref{fig:LyaKinematicsALL}), ensuring that all of the Ly$\alpha$ velocities are included in flux measurement.

Integrating over the region encompassed by the over-plotted MOSFIRE N1 slit within the aforementioned spatial window of $\sim$81.76 kpc, results in a total Ly$\alpha$ flux of $\rm{F_{Ly\alpha}}=1.44\pm0.10\times 10^{-16}$ erg/cm$^2$/s (equivalent to a surface brightness of $\rm{SB_{Ly\alpha}}=1.48\pm0.10\times 10^{-17}$ erg/cm$^2$/s/\sq\arcsec).

\subsection{The H$\alpha$ Emission of the Slug Nebula}
\label{sec:Haemission}

\begin{figure}
\begin{center}
\includegraphics[width=0.5\textwidth]{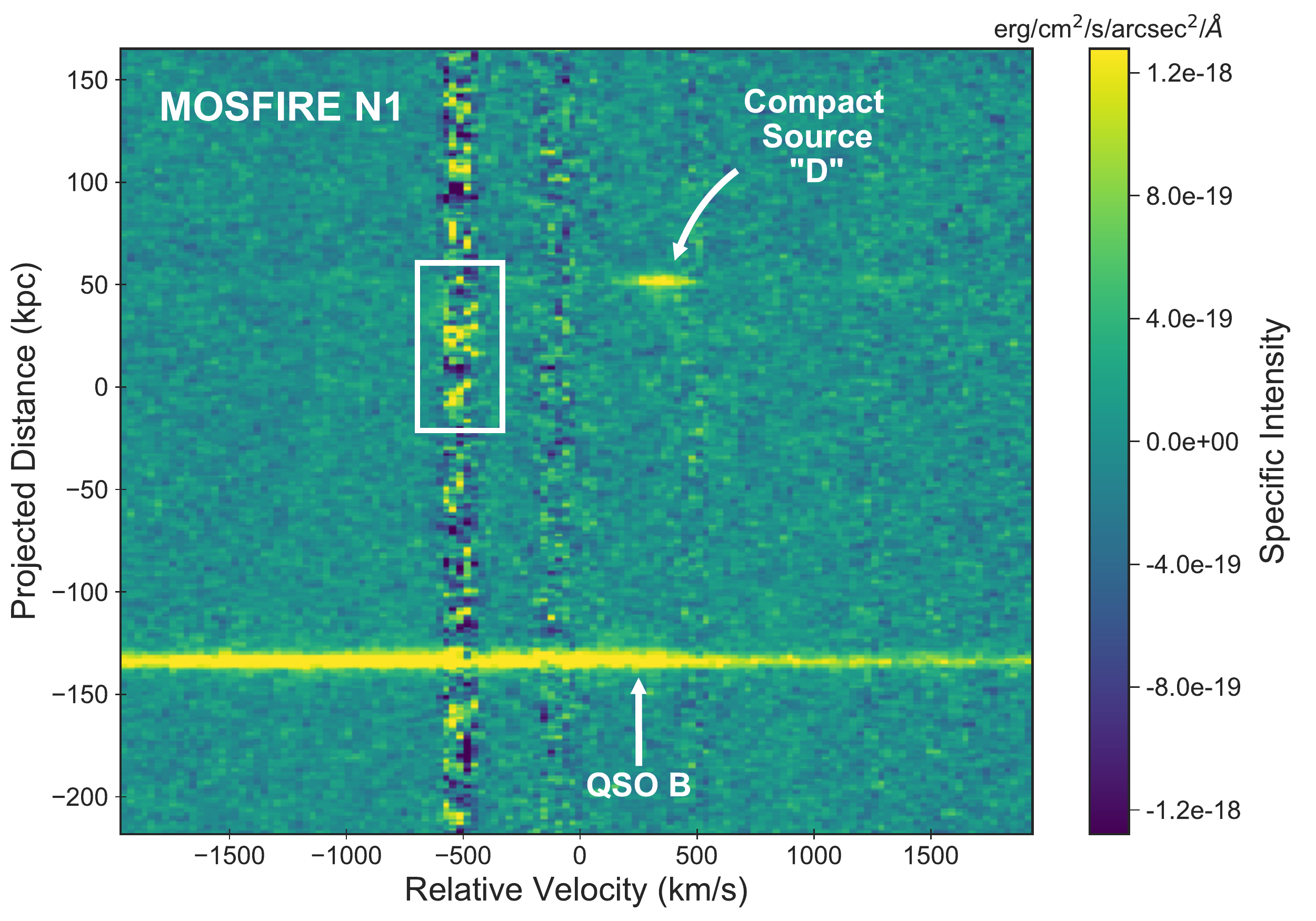}
\includegraphics[width=0.5\textwidth]{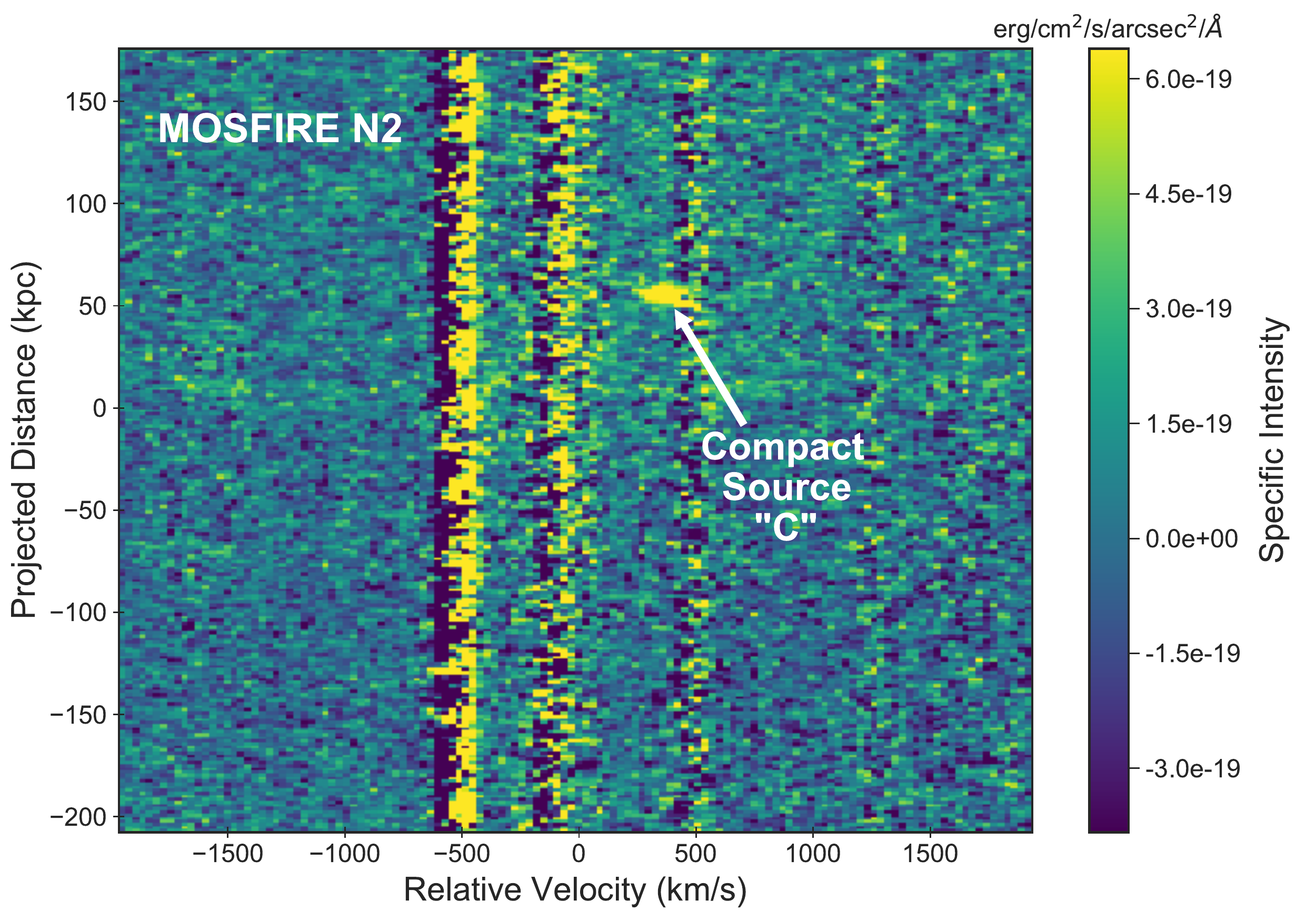}
\end{center}
\caption{The top panel shows the unsmoothed two-dimensional spectrum taken with Keck I/MOSFIRE (4.8 hrs) using the N1 slit orientation (the red slit in Figure \ref{fig:SlitPA}). The $v=0$ km/s corresponds to the expected H$\alpha$ emission at a redshift of z=2.283, the redshift of QSOA. The spatial offset of 0 kpc indicates the location of the MOSFIRE N1 and LRIS slit intersection, referred to as position ``X" in Figure \ref{fig:SlitPA}. The white rectangle indicates the region in which the H$\alpha$ flux of the Slug was measured (see \S \ref{sec:Haaperture} for a description of how the dimensions and location of the rectangle were chosen). 
Note that the flux measurement aperture overlaps with the continuum emission from compact source ``D''. This contaminant is masked out when we perform any analyses or measure fluxes. The bright continuum source around $\sim$150 kpc is QSO B. The H$\alpha$ and \ion{N}{II}[6583] emission lines of compact source ``D'' (z=2.287) are visible at a spatial position of $\sim 50$ kpc and spectral positions of $\sim$ 400 km/s and $\sim$ 1200 km/s respectively.  The bottom panel shows the unsmoothed two-dimensional spectrum taken with Keck I/MOSFIRE (2.6 hrs) using the N2 slit orientation (the green slit in Figure \ref{fig:SlitPA}). The spectral and spatial axes match those of the N1 spectrum and their zero-points are defined in the same way as in the top panel. The emission line at $\sim 400 \rm{km/s}$ and $\sim 50 \rm{kpc}$ is the H$\alpha$ line of compact source ``C'' (z=2.287).}
\label{fig:Mosfire2Dspec}
\end{figure}

\begin{figure}
\begin{center}
\includegraphics[width=0.5\textwidth]{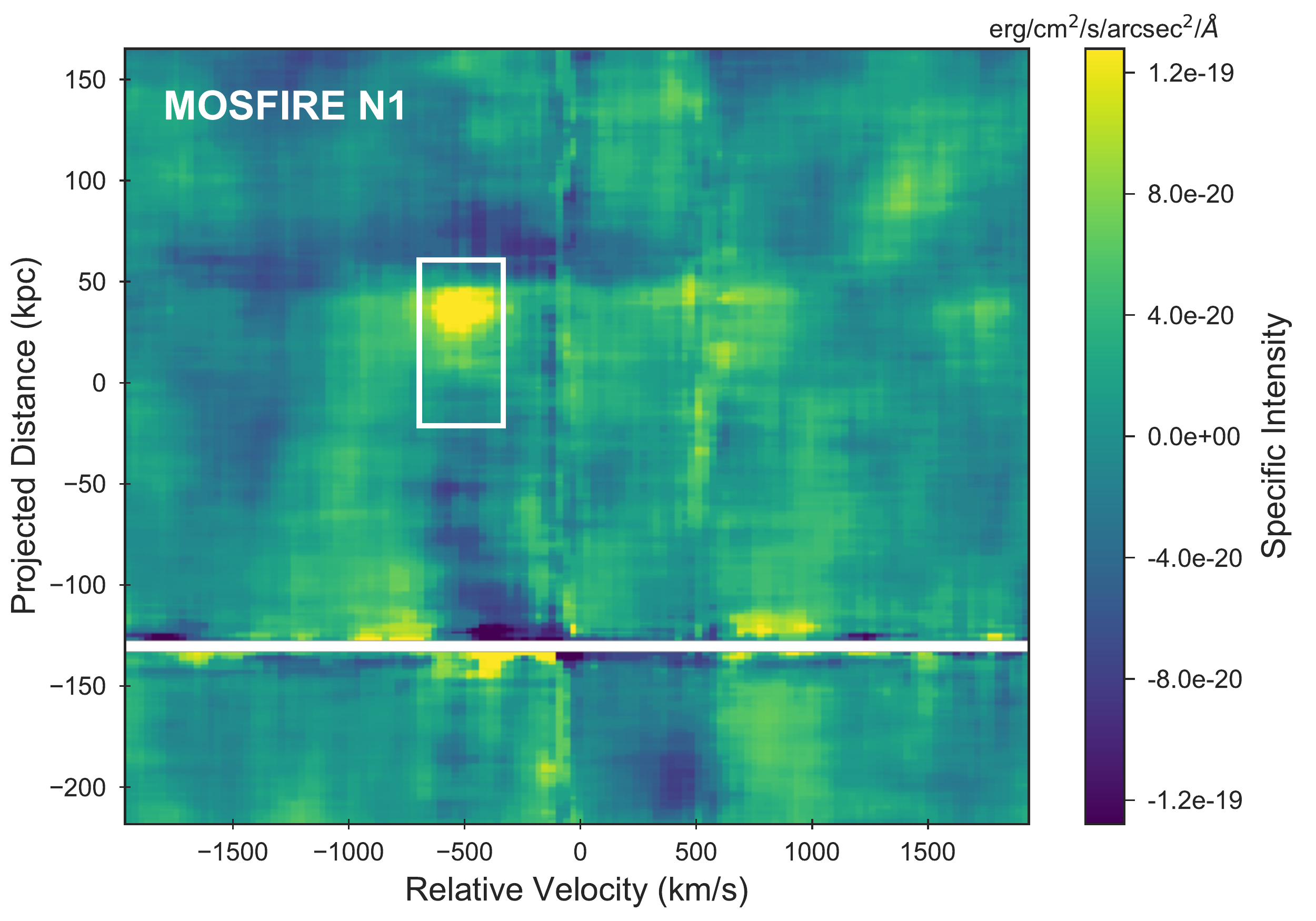}
\end{center}
\caption{The smoothed MOSFIRE N1 2D spectrum produced using a median filter with a smoothing kernel of 41 kpc (27 pixels) in the spatial direction and 363 km/s (12 pixels) in the spectral direction. These dimensions correspond to half the size of the white rectangular aperture that was used to measure the H$\alpha$ flux.  Prior to smoothing, the continuum and line emission from compact source ``D'' as well as QSO B were masked out. Since the median-smoothing filter does not conserve flux, this figure is meant to be purely illustrative and was not used for any of the measurements in the analysis. Though the H$\alpha$ detection lies on top of a relatively bright sky line, we argue in \S \ref{sec:therealHa} that the emission is produced by the Slug Nebula rather than high variance pixels in the sky line residual.} 
\label{fig:Mosfire2Dspec_smoothed}
\end{figure}

\subsubsection{Determining the optimal aperture for H$\alpha$ detection and flux measurement}
\label{sec:Haaperture}

\begin{figure}
\begin{center}

\includegraphics[width=0.5\textwidth]{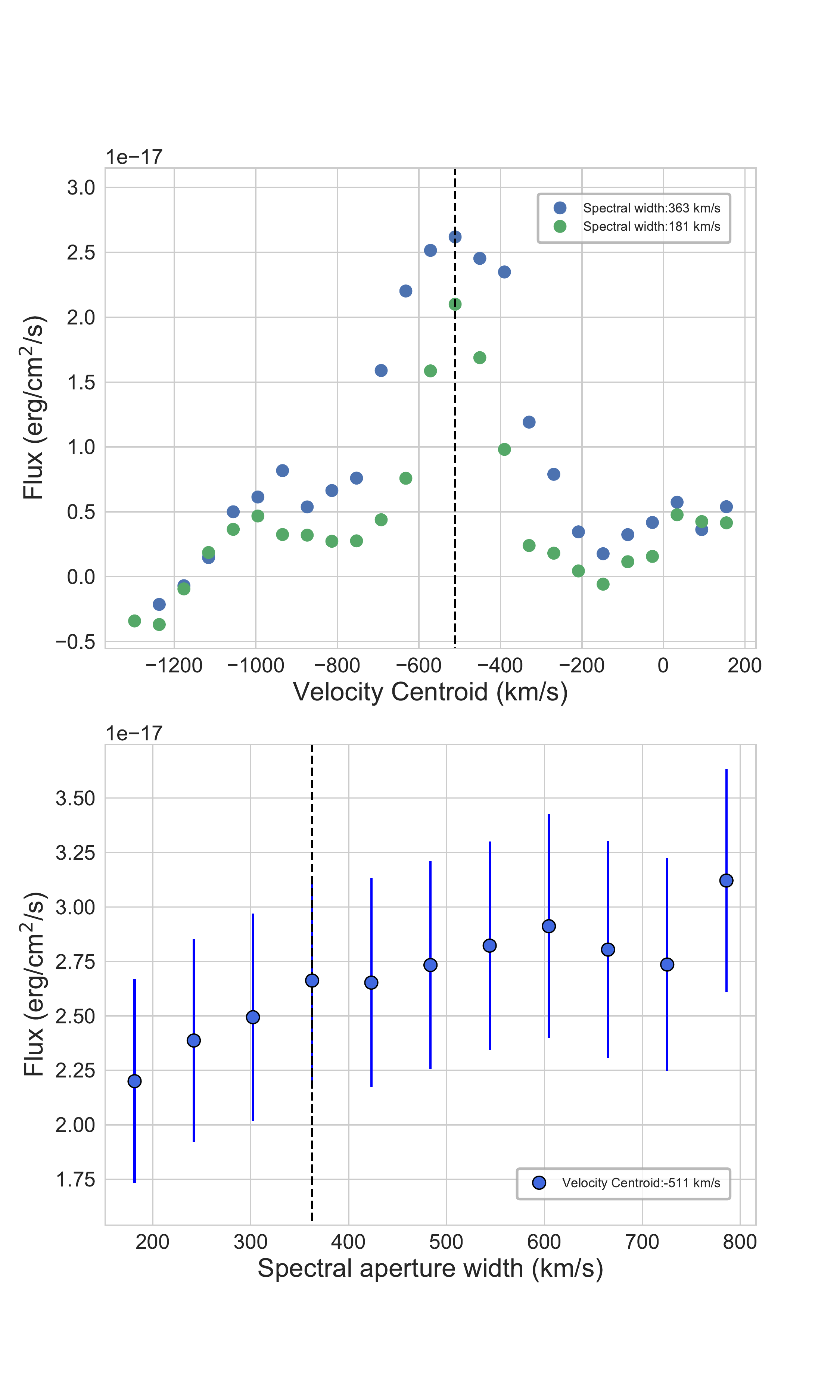}
\end{center}
\caption{Though the spatial aperture size and centroid can be determined empirically from the narrow-band Ly$\alpha$ image (see Figure \ref{fig:SlitPA}), radiative transfer effects can modify the kinematics of the Ly$\alpha$ such that it cannot be used to inform the expected wavelength of the H$\alpha$ emission. To determine the most likely velocity centroid, we calculated the flux as a function of velocity position within a narrow velocity window (6 pixels/ 181 km/s), as shown in green in the top panel of this figure. The velocity that maximized the flux, $-$511 km/s, was chosen as the centroid position. Note that doubling the velocity window does not change the location of the peak (shown in blue). We used a flux curve of growth approach to determine the optimal spectral window size. The flux within the MOSFIRE aperture (centered at the velocity centroid of -511 km/s) is shown in the bottom panel as a function of velocity aperture width. At a spectral width of 363 km/s, the H$\alpha$ flux starts to level off. We therefore choose this as the size of the velocity width of our aperture when calculating the Slug's H$\alpha$ flux. The flux within these aperture dimensions result in a SNR of $\sim 5.6 \sigma$. }
\label{fig:MOSanalysisvelWG}
\end{figure}

The large spatial scale (and possibly, the large velocity width) of the expected H$\alpha$ emission from the Slug Nebula necessarily requires
spatial and spectral binning of our original data presented in Figure \ref{fig:Mosfire2Dspec}. Moreover, we do not know a priori where the spatial and velocity center
of such an aperture should be located. 
 
In this section, we discuss how we obtained the optimal rectangular aperture for the detection of the H$\alpha$ emission from the Slug. Because of the lower exposure time and higher systematic noise of the MOSFIRE Night 2 observations (see Figure  \ref{fig:Mosfire2Dspec}) we will limit our search for and analysis of extended emission to the MOSFIRE Night 1 observations here
and in the remainder of the paper. However, we will make use of the MOSFIRE Night 2 observations for the spectral analysis of the compact source ``C".

As discussed in this section, we find that the optimal aperture has a spatial dimension of $\sim$81.76 kpc and is centered a distance of $\sim$19.68 kpc from the intersection of 
the MOSFIRE Night 1 and Night 2 slits. The optimal spectral dimension has a width of 363 km/s centered at a velocity of -511 km/s with respect 
to the systemic velocity of H$\alpha$ at a redshift of z=2.283 (the systemic redshift of the quasar UM287 obtained from CO observation; Decarli et al, in prep.). 

The spatial scale and center were chosen based on the intersection of the Night 1 slit with the Ly$\alpha$ NB emission. The narrow band Ly$\alpha$ surface brightness was first rescaled to an expected H$\alpha$ surface brightness assuming a case B recombination ratio of $SB_{\rm{Ly\alpha}} / SB_{\rm{H\alpha}}=8.1$. We then calculated a 3$\sigma$ contour, assuming an estimated MOSFIRE H$\alpha$ surface brightness error of $1\sigma = 3.7\times 10^{-19} ~\rm{erg/cm^2/arcsec^2/s}$. The region within the intersections of the Night 1 slit and the 3$\sigma$ contour was 9.71$\arcsec$ long and centered a distance of 2.34$\arcsec$ from the intersection of the Night 1 and Night 2 slits. Assuming a redshift of z=2.283, this translates to a spatial aperture in which to calculate the H$\alpha$ flux of $\sim$81.76 kpc centered at a distance of $\sim$19.68 kpc from the slit intersection.

We then determined the optimal spectral aperture width and central velocity. In the absence of radiative transfer effects influencing the velocity distribution of the Ly$\alpha$ emission, we would expect that the H$\alpha$ emission to be centered close to the velocity centroid of the Ly$\alpha$, found to be $-555$ km/s in \S \ref{sec:LRISkinematics}. However, possible asymmetries in the Ly$\alpha$ due to scattering effects could bias our determination of the precise H$\alpha$ central velocity. 

In order to allow for this possibility, we chose a ``priorless" approach to determining the velocity centroid and width of the H$\alpha$ emission. Rather than select the velocity centroid based on it's expected location, we took a curve of growth approach, finding the central velocity in a wide velocity window (shown in the top panel of Figure \ref{fig:MOSanalysisvelWG}), that maximized the H$\alpha$ flux. We first masked out the compact source continuum emission located at $\sim$50 kpc from the slit intersection ``X", then the H$\alpha$ flux was measured assuming a narrow velocity window of 181 km/s (6 pixels) so as to finely sample the velocity range. The H$\alpha$ flux peaks at a velocity centroid of -511 km/s, a result that is corroborated if we double the velocity width to 363 km/s, as shown in the top panel of Figure \ref{fig:MOSanalysisvelWG}. The curve-of-growth determined velocity centroid of -511 km/s is extremely close to the velocity centroid of the Ly$\alpha$ region 2 emission, which was measured to be -555 km/s. 

Finally, we determined the spectral aperture width, which cannot simply be obtained from the breadth of the Ly$\alpha$ emission, since radiative effects can broaden the width of this resonant line. Instead, we varied the spectral aperture width from 181 km/s to 784 km/s and selected the width at which the measured H$\alpha$ flux leveled off. As shown in the bottom panel of Figure \ref{fig:MOSanalysisvelWG}, the optimal velocity aperture has a width of 363 km/s. 

Since this curve of growth approach to finding the velocity centroid and width of the flux aperture seeks to maximize the H$\alpha$ flux, one could be concerned that this approach would consistently bias our H$\alpha$ flux towards higher values. In order to quantify this effect, we used the same methodology described above to find the peak H$\alpha$ flux in several pure sky background regions. When we varied the velocity centroid and width of the flux apertures, we consistently found that the maximum H$\alpha$ peaks exceeded the mean flux value in that sky background region by about 5$\times 10^{-18}$ erg/cm$^2$/s. Therefore, we conclude that this ``priorless" aperture selection would inflate the H$\alpha$ flux by $\leq 5 \times 10^{-18}$ erg/cm$^2$/s. We emphasize that our estimate is an upper limit, as it is unlikely that a statistical fluctuation would land near or on top of the detected H$\alpha$ flux.

\subsubsection{An empirical estimate of the sky noise}
\label{sec:Haerror}

We determined an empirical noise estimate by calculating the standard deviation of the flux in ``pure-sky" regions. These regions were chosen so as to avoid the expected spatial location of Slug H$\alpha$ emission as well as the outer edges of the slit, which have a reduced total exposure time. We measured the flux in these background regions using the same-sized rectangular aperture and velocity centroid as when measuring the Slug Nebula flux (see \S \ref{sec:Halphaflux}). We found that these fluxes were dominated by a linearly varying, spatially-dependent background gradient, which we modeled and removed prior to calculating the flux scatter for the pure-sky regions. We note that the removal of this background model does not affect our measurement of the Slug's H$\alpha$ flux since the estimated background was very close to zero at that spatial position. 

\begin{figure}
\begin{center}
\includegraphics[width=0.5\textwidth]{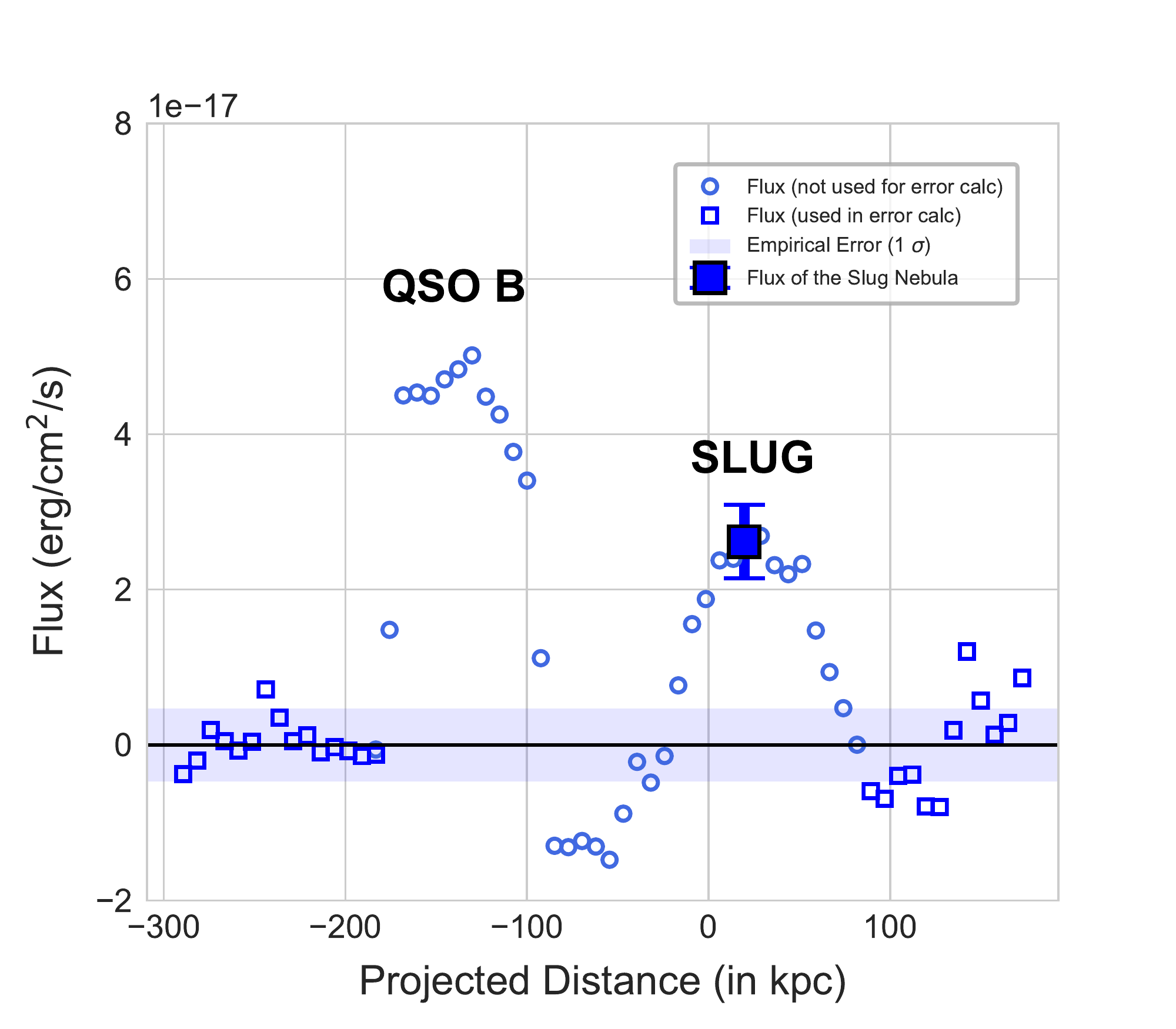}
\end{center}
\caption{The H$\alpha$ flux computed within rectangular apertures of size 81.76 kpc by 363 km/s, measured as a function of spatial position along the MOSFIRE N1 slit. For all these flux calculations, the apertures were centered at a velocity of -511 km/s. The bigger filled blue square point marks the flux at the expected location of the Slug Nebula (see Figure \ref{fig:Mosfire2Dspec}). The light blue points with much higher flux values ($\sim$ -75 kpc to $\sim$75 kpc) are associated with the QSOB emission. The small darker blue unfilled squares correspond to regions far enough away from the expected location of the Slug Nebula to be considered ``pure sky". The error on the measurement of the Slug Nebula flux was calculated using the standard deviation of the flux in these ``pure sky" apertures. We note here that the ``pure sky'' regions have a background gradient removed but the rest of data do not, see \S \ref{sec:Haerror} for details. The $\pm1\sigma$  error is shown as the transparent blue region as well as the blue error bars associated with the Slug Nebula flux (large blue square). The flux at the location of the Slug Nebula (the dark blue square) corresponds to a $\sim 5.6 \sigma$ detection. }
\label{fig:MosfireHaFlux}
\end{figure}

The ``pure sky'' fluxes, with the background gradient removed, are plotted in Figure \ref{fig:MosfireHaFlux} as unfilled blue squares. The $\pm1 \sigma$ standard deviation of the sky fluxes (our empirical noise estimate) is shown as the transparent blue shaded region. The light blue unfilled circles show the flux at spatial locations close to the expected H$\alpha$ emission and were not included in the calculation of our noise estimate. The background gradient was not removed at the location of the light blue unfilled circles.  The flux at the location of the Slug is represented by the larger filled blue square. 

\subsection{Examining the Robustness of the Slug Nebula H$\alpha$ Detection}
\label{sec:HaRobust}

\begin{figure*}
\centering
\hspace*{\fill}%
\subfloat[]{\includegraphics[width=0.5\textwidth]{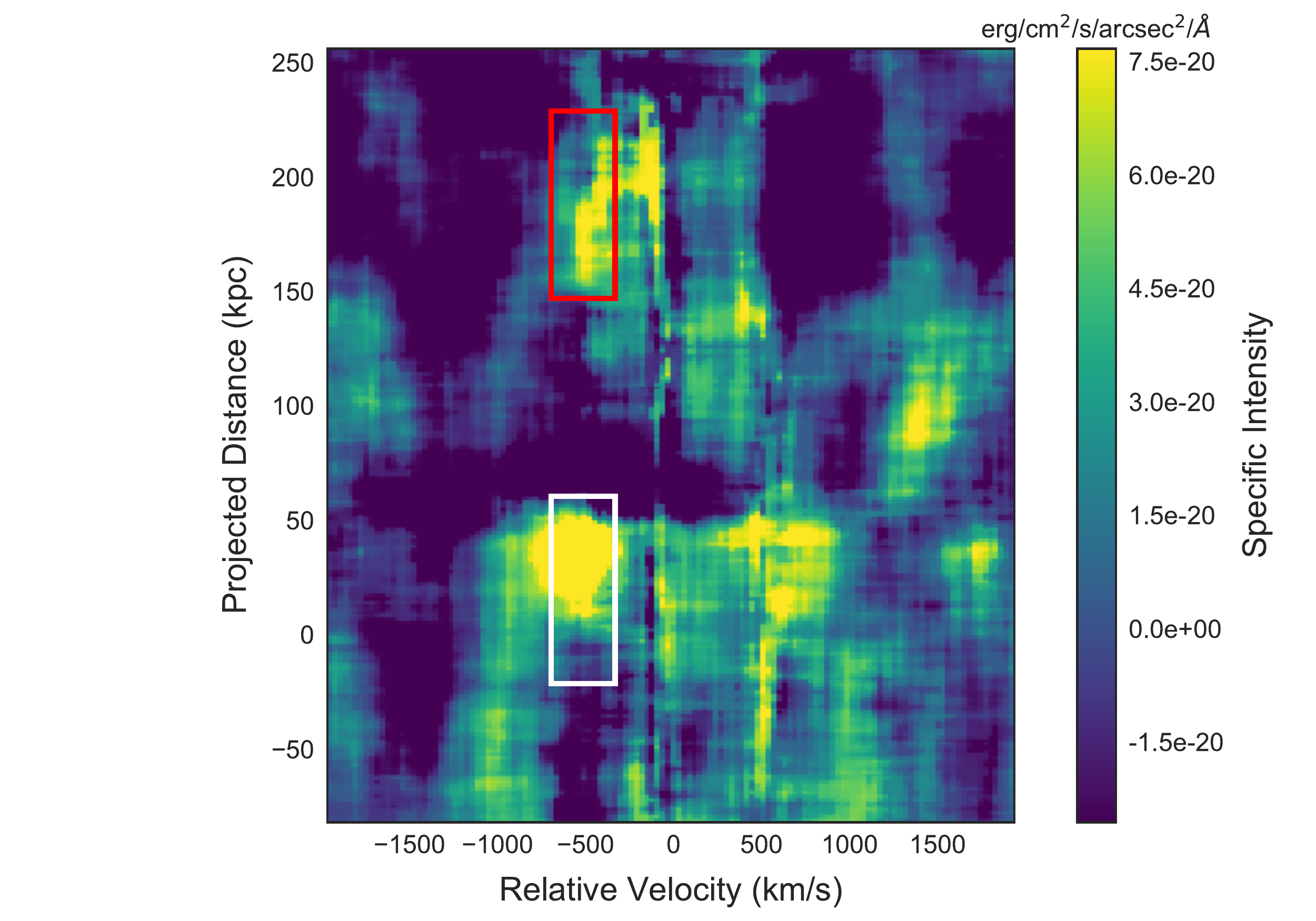}}%
\hfill
\subfloat[]{\includegraphics[width=0.5\textwidth]{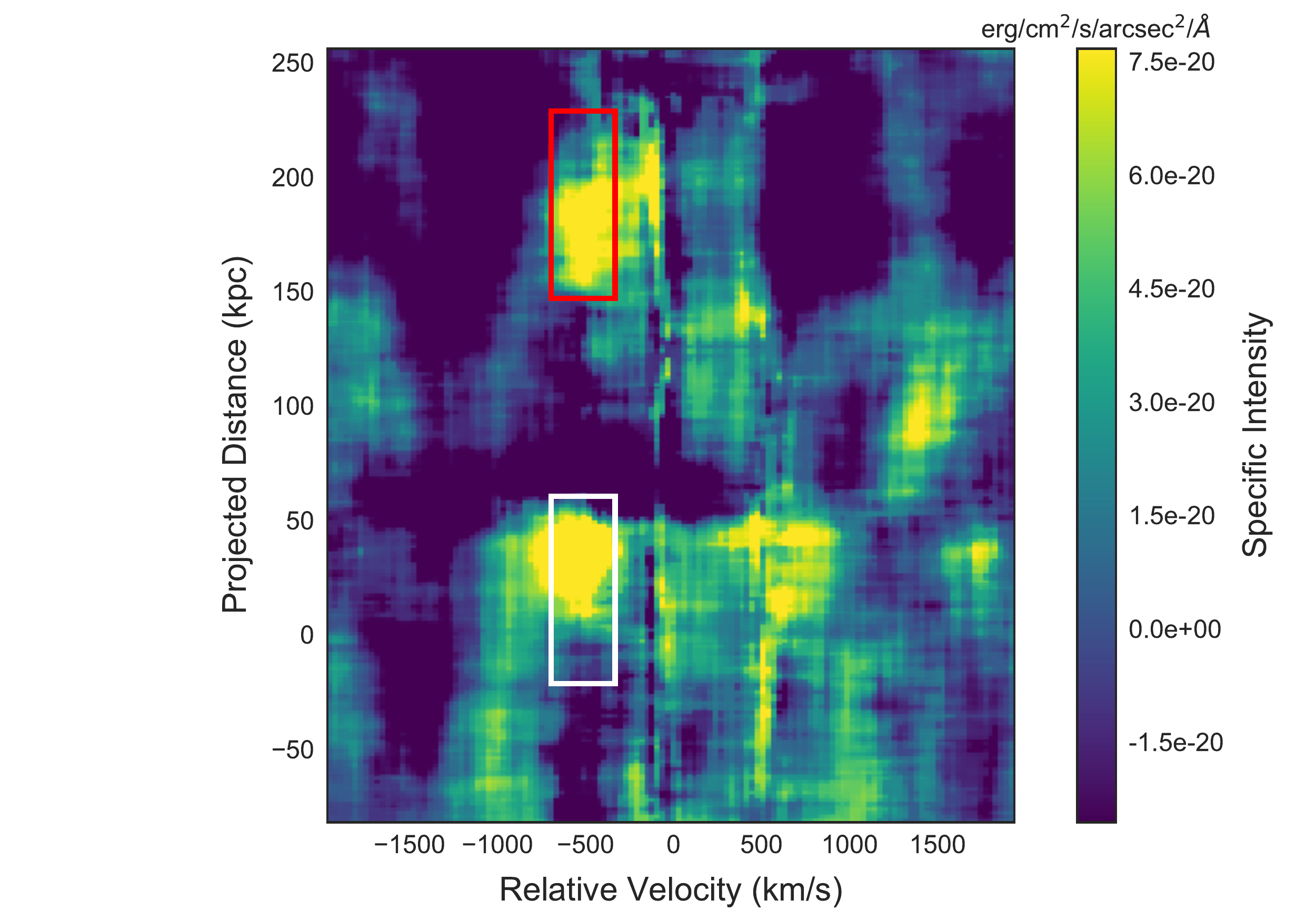}}%
\hfill

\caption{ This figure presents a visual verification that our observed H$\alpha$ flux (in the white rectangle) resembles a reasonable model of the H$\alpha$ emission of the Slug. The injected fake sources represent two extremes of the possible H$\alpha$ velocity distributions: a) the H$\alpha$ emission is assumed to have a broad velocity distribution that matches that of the LRIS Ly$\alpha$ spectrum (left panel) and b) the H$\alpha$ line is assumed to have been emitted at a much more narrow range of velocities and the Ly$\alpha$ was broadened by radiative transfer effects (right panel). 
The left panel shows the median-smoothed image of the MOSFIRE N1 spectrum (using a kernel of 41 kpc x 363 km/s) with an injected fake source that was modeled by taking the Ly$\alpha$ flux from ``region 2" in the LRIS spectrum and rescaling it to match the H$\alpha$ flux. Thus, this H$\alpha$ emission model keeps the velocity distribution of the Ly$\alpha$ emission intact and is centered at a spatial position of 188 kpc and at a velocity of -718 km/s. We find that the brightest region of the injected source (expected location shown as the red rectangle) appears less significant than the observed emission (in the white rectangle). Similarly, in the right panel we show the MOSFIRE N1 spectrum median-smoothed with the same kernel of 41 kpc $\times$ 363 km/s that was applied in Figure \ref{fig:Mosfire2Dspec_smoothed}.  The injected fake source, centered at 188 kpc and a velocity centroid that matches the observed H$\alpha$ emission, was modeled as a 2D-Gaussian. The fake emission had a standard deviation in the spatial direction of 18 kpc and of 181 km/s in the spectral direction and a total flux of $2.62\times10^{-17}$ erg/s/cm$^2$ that matches the observed H$\alpha$ flux. This emission model (visible within the red rectangle) better resembles the observed H$\alpha$ emission (at the location of the white rectangle), suggesting that the H$\alpha$ emission of the Slug has different kinematics than the Ly$\alpha$ and is emitted at a narrower range of velocities.} 
\label{fig:FakeSources}
\end{figure*}

It is important to note that our chosen velocity centroid is coincident with the sky line at 20517\AA. The presence of a bright, imperfectly subtracted sky line at the location of our H$\alpha$ detection might cause concern that the observed H$\alpha$ flux within our chosen aperture is due to variance in the sky line rather than emission from the Slug Nebula. 

Figure \ref{fig:Mosfire2Dspec} shows no clear emission at the location of the chosen aperture. However, once the spectrum is smoothed using a median filter with a 41 kpc $\times$ 363 km/s kernel, the H$\alpha$ emission becomes apparent, as seen within the white rectangle in the top panel of Figure \ref{fig:Mosfire2Dspec_smoothed}.  In addition, this sort of line-like emission appears nowhere else along the skyline or generally in the vicinity of the expected H$\alpha$ emission.  

The idea that the emission within our aperture is uncharacteristic of the variance of the sky line is corroborated by the fact that the signal to noise also peaks at the same velocity centroid as the flux.  The error used in the SNR was empirically calculated by taking the standard deviation of the flux in apertures along the sky line (see \S \ref{sec:Haerror}). Therefore, if the emission in our aperture was typical of the sky line, this would be reflected in the noise estimate. While the flux could be biased by the presence of a sky line, the signal to noise ratio should be much less susceptible to this effect. The velocity centroid corresponding to the peak SNR was unchanged whether we used our empirical noise estimate or a noise estimate calculated from the error array produced by the MOSFIRE DRP. 

As an additional test of the validity of the H$\alpha$ emission, we inserted two types of fake sources into the MOSFIRE N1 spectrum in order to verify that observed emission line is consistent with what would be predicted from the Ly$\alpha$.  The first fake source was created by taking the Ly$\alpha$ emission within region 2 of the LRIS slit and rescaling it to the H$\alpha$ flux ($2.62\times10^{-17}$ erg/cm$^2$/s).
It was then inserted it into the MOSFIRE N1 spectrum at a velocity window matching that of the Ly$\alpha$ kinematics (centered at -718 km/s) but at a spatial location away from the expected Slug emission. The results are shown in the left panel of Figure \ref{fig:FakeSources}, with the red rectangle denoting the fake source and the white rectangle the actual observed emission at the location of the Slug. 

Since the Ly$\alpha$ could be broadened by radiative transfer effects that would not affect the H$\alpha$ emission, the H$\alpha$ could be emitted with a much more concentrated velocity distribution.  The second fake source, inserted into the MOSFIRE N1 spectrum at the observed H$\alpha$ velocity centroid (-511 km/s), was chosen to be a 2D gaussian with $\sigma_{\rm{vel}}$=181 km/s and $\sigma_{\rm{spat}}$=18 kpc and a total flux equivalent to that of the detected H$\alpha$ flux. 

As seen in the right panel of Figure \ref{fig:FakeSources}, the observed H$\alpha$ emission (white rectangle) looks similar to the compact 2D gaussian fake source (red rectangle). As this exercise was purely for illustrative purposes, the fact that the observed emission looks similar to a reasonable expectation of the H$\alpha$ emission supports that the observed H$\alpha$ emission is not simply due to the underlying sky line. In addition, the apparent compact size of the H$\alpha$ as compared with the expected size seen in Ly$\alpha$ could suggest that the Ly$\alpha$ emission is broaden by radiative transfer effects. 

\subsubsection{The H$\alpha$ flux and the Ly$\alpha$ to H$\alpha$ ratio}
\label{sec:Halphaflux}

We measured the H$\alpha$ flux of the observed portion of the Slug Nebula in the region of the MOSFIRE N1 slit using the rectangular aperture obtained as discussed above. This aperture has a spatial dimension of 81.76 kpc and spectral dimension of 363 km/s and it is spatially centered at a distance of 19.68 kpc from the intersection of the Night 1 and Night 2 slits (point ``X" in Figure \ref{fig:SlitPA}). The velocity centroid of the aperture is -511 km/s. The region in which the H$\alpha$ flux was measured is over-plotted as a white rectangle in the top panel of Figure \ref{fig:Mosfire2Dspec} and both panels of Figure \ref{fig:Mosfire2Dspec_smoothed}. 

We find an H$\alpha$ flux within our aperture of F$_{\rm{H\alpha}}= 2.62\pm 0.47 \times 10^{-17}$ erg/cm$^2$/s  (equivalent to a surface brightness of SB$_{\rm{H\alpha}}=2.70\pm 0.48 \times 10^{-18}$ erg/cm$^2$/s/\sq\arcsec), where the error is calculated from the standard deviation of the fluxes in ``pure sky" regions as described in \S \ref{sec:Haaperture}. Considering the Ly$\alpha$ flux in the same spatial region obtained from the NB image (found to be $\rm{F_{Ly\alpha}}=1.44\pm0.10\times 10^{-16}$ erg/cm$^2$/s in \S \ref{sec:Lyaflux}), the Ly$\alpha$ to H$\alpha$ flux ratio in this region of the Slug is 5.5$\pm$1.1. If, as discussed in \S \ref{sec:Haaperture}, we take into account that the H$\alpha$ flux might be biased high by up to $5\times 10^{-18}$ erg/cm$^2$/s, the Ly$\alpha$ to H$\alpha$ flux ratio would instead be around 6.9. We will discuss the possible implications of this flux ratio with respect to the physical emission mechanism and Ly$\alpha$ escape fraction in \S \ref{sec:fluorescentnature}.

\subsection{The Compact Sources in the MOSFIRE Data }
\label{sec:compactsources}
Two line emitters were also observed in our MOSFIRE spectra. These sources were originally detected in the LRIS NB and V band data and were dubbed compact source ``D" and compact source ``C". The sources are shown and labeled in Figure \ref{fig:SlitPA} and correspond to the MOSFIRE N1 and MOSFIRE N2 slits respectively. Note that emitter ``C" is the same as source ``C" in \citet{Martin2015} and the compact source in our LRIS spectrum (see Figure \ref{fig:LRIS2Dspec}). 

The bright H$\alpha$ emission line of sources ``D" and ``C" in our MOSFIRE spectra allow us to determine the redshifts of each compact source. We find that both sources are at the same redshift of z$\approx 2.287$, which is slightly redshifted with respect to that of QSO A. This corresponds to a velocity offset between the compact sources and QSO A of {$\sim$355} km s$^{-1}$ that could be explained by peculiar motions within the halo of QSO A. 

In addition to computing the redshifts, we also produced a K-band 1D spectrum for each source using a simple box extraction. These spectra were used to calculate the H$\alpha$ and corresponding \ion{N}{II}[6583] line fluxes. For compact source ``D", we find an H$\alpha$ flux of F$_{\rm{H\alpha}}=$6.6$\pm 0.3\times 10^{-17}$ erg/cm$^2$/s and an \ion{N}{II} flux of F$_{\rm{\ion{N}{II}}}=$2.2$\pm 0.2\times 10^{-17}$ erg/cm$^2$/s. For compact source ``C", we find an H$\alpha$ flux of F$_{\rm{H\alpha}}=$4.3$\pm 0.4\times 10^{-17}$ erg/cm$^2$/s and a \ion{N}{II} 3$\sigma$ flux upper-limit of F$_{\rm{\ion{N}{II}}}=$2.4$\times 10^{-18}$ erg/cm$^2$/s.

\begin{figure}
\begin{center}
\includegraphics[width=0.5\textwidth]{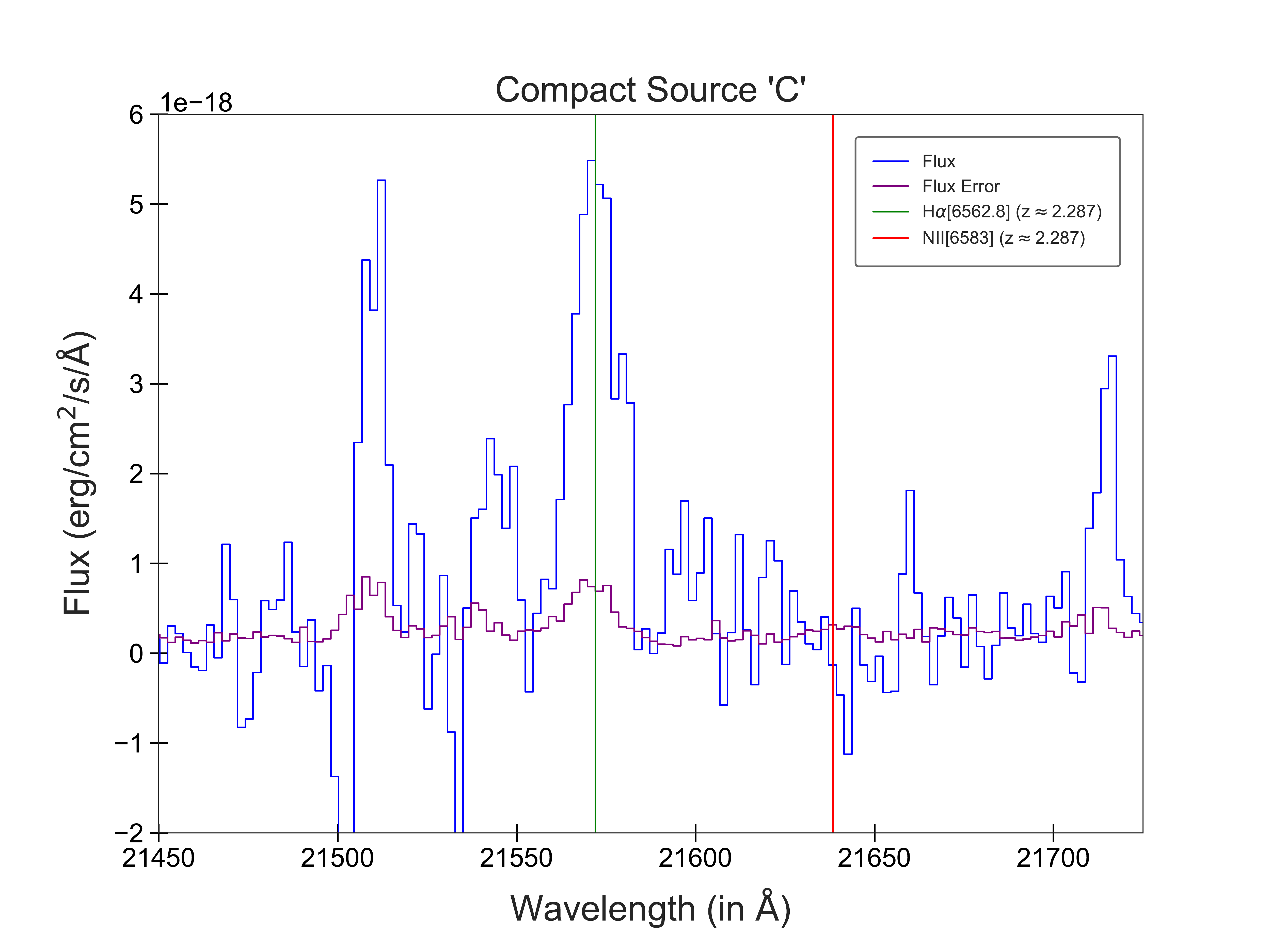}
\includegraphics[width=0.5\textwidth]{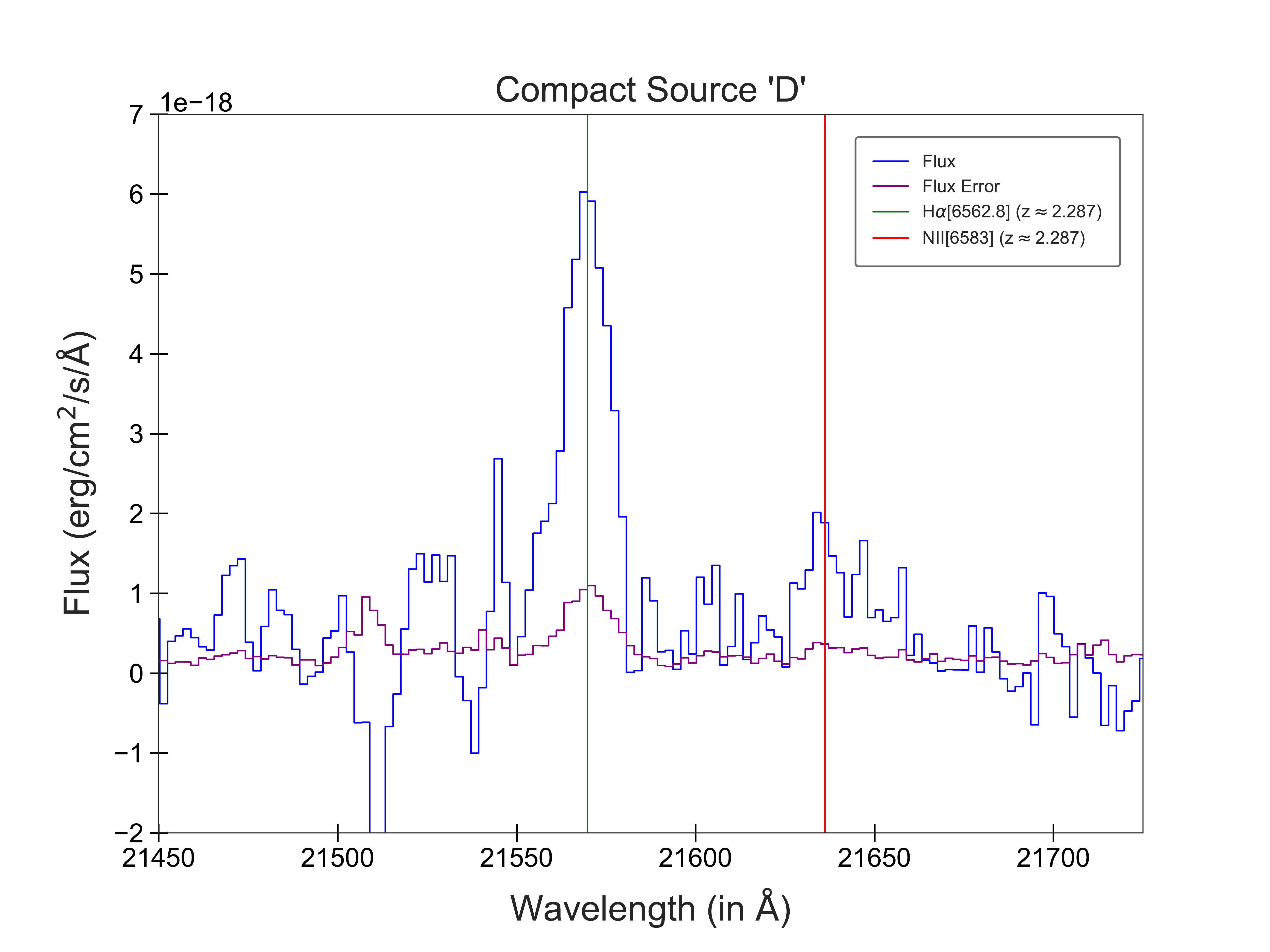}
\end{center}
\caption{The 1D spectra of compact sources ``C" (top panel) and ``D" (bottom panel) derived from the 2D MOSFIRE N2 and N1 spectra, respectively, using a simple box extraction. The flux is shown in blue, the error array is plotted in purple, while the green and red vertical lines shows the expected wavelength of H$\alpha$ and \ion{N}{II}[6583] respectively, given the redshift of z=2.287 for each source.}
\label{fig:CompactSourceSpec}
\end{figure}

\section{Discussion}
In this work, we presented deep spectroscopy of the Ly$\alpha$ and H$\alpha$ emission of the Slug Nebula. We use these data in the following sections to discuss the physical structure of the Slug Nebula's gas as well as the production mechanism of the Slug's Ly$\alpha$ emission and its implications. In \S \ref{sec:DiscussLyaKinematics}, we use the kinematic indicators derived from the Keck/LRIS spectroscopy (see \S \ref{sec:LRISkinematics}) to address the question of how the Slug's gas is physically distributed. 

In \S \ref{sec:emissionmech}, the Slug's H$\alpha$ flux, measured in \S \ref{sec:Halphaflux}, is compared to it's corresponding Ly$\alpha$ flux (\S \ref{sec:Lyaflux}) so as to determine the emission mechanism producing the Slug's observed Ly$\alpha$. In particular, we discriminate between two scenarios for the production of the Ly$\alpha$ emission: 1) a purely ex-situ production of the Ly$\alpha$ in the broad-line region of QSO A that is then scattered and reemitted by neutral hydrogen in the nebula and 2) a significant contribution of in-situ fluorescent Ly$\alpha$ emission produced by case B recombination of the Slug's hydrogen gas.  

Finally, we also examine the origin of the H$\alpha$ emission of compact sources ``C" and ``D".  In \S \ref{sec:DiscussCompact}, we use the ratio of their \ion{N}{II}/H$\alpha$ flux to place these galaxies on a BPT diagram and determine whether these galaxies are star-forming or have a central AGN. In addition, we also explore the possibility of a contribution of fluorescent emission due to QSO A.

\subsection{The Ly$\alpha$ kinematics of the Nebula}
\label{sec:DiscussLyaKinematics}

We can gain insight into the physical structure of the gas by examining the Ly$\alpha$ kinematics. In their work,  \citet{Martin2015} claimed that the brightest region of the Slug Nebula is an extended rotating hydrogen disk contained within an $\approx10^{13}\rm{M}_{\odot}$ dark matter halo. However, the kinematics shown in Figure \ref{fig:LyaKinematicsALL} belie the idea that the Slug Nebula is a simple monolithic structure like a disk. 

As discussed in \S \ref{sec:LRISkinematics}, the velocity centroid as a function of spatial position (the upper-left panel of Figure \ref{fig:LyaKinematicsALL}) reveals three clearly distinguishable regions with distinct velocity centroids. These same regions are clearly recognizable in the plot of velocity dispersion as a function of spatial position and are marked by very sharp transitions at $\sim -35$ kpc and $\sim 45$ kpc. 

The two left-most regions comprise the Slug Nebula. The dimmer ``region 1" is centered at $\sim -50$ kpc with a characteristic velocity centroid of -333 km/s while the brighter ``region 2" is located at $\sim 25$ kpc with a velocity centroid of -555 km/s. The Ly$\alpha$ emission of the Slug is separated from that of compact source ``C" by a very narrow transition region ($\sim$10 kpc). We also see that the velocity dispersion has sharp transitions at the same location where we see sharp changes in the mean velocity, lending further credence to our interpretation that these are kinematically distinct regions.

Although it is difficult to disentangle velocity effects from distances along the line of sight, these sharp transitions may suggest
that the Slug Nebula could be composed of several structures. This is not unexpected from our theoretical understanding of 
cosmic structure formation: the most massive filaments of the cosmic web are composed of both diffuse material and more massive
haloes containing denser gas. If the Ly$\alpha$ emission is produced by recombination radiation and therefore scales with the gas density
squared, our observations would be most sensitive to detecting the densest \emph{knots} and structures within the filaments.

This interpretation of the Slug's physical structure is inconsistent with the giant disk argued for by \citet{Martin2015}, despite the fact that \citeauthor{Martin2015}'s pseudo-slit largely overlaps with our LRIS slit.  We believe that the lower spatial and spectral resolution of the pseudo slit observations may have smoothed out the sharp transitions that we resolve, making the distribution of velocity centroids resemble that of a giant disk. 

Instead, our observation reveals a very abrupt cutoff, seen in Figure \ref{fig:LRIS2Dspec}, of the Ly$\alpha$ flux at a spatial position of $\sim 50$kpc. It is currently unclear whether this sharp edge to the Ly$\alpha$ emission is due to an absence of cold gas or whether the Ly$\alpha$ emission is instead being absorbed along the line of sight. The cutoff location is tantalizingly close to compact source ``C". However, compact source ``C" is located several hundred km/s on the red side of this feature and therefore we do not think that the compact source and its environment are the origin of a possible absorption feature. It is also interesting to note that the Ly$\alpha$ emission cutoff is adjacent to the brightest region of the nebular emission. Deeper integral field observations with Keck/KCWI and MUSE for both Ly$\alpha$ and other emission lines (Cantalupo et al, in prep.) could be useful to disentangle whether this may be due to a lack of quasar illumination in a particular direction (i.e., a ``shadow" of some absorber that is associated with the quasar) or suggestive of a possible different physical origin for this emission (e.g, shocks).

\subsection{Constraining the Emission Mechanism of the Slug Nebula}
\label{sec:emissionmech}

\citet{Cantalupo2014} presents two possible mechanisms that could power the Ly$\alpha$ emission of the Slug Nebula. In the first case, Lyman continuum photons produced by the nearby QSO A ionize the gas of the nebula, producing Ly$\alpha$ photons as the hydrogen atoms recombine. In the so-called ``case A", the gas is optically thin to ionizing radiation, while the opposite optically thick situation is usually referred to as ``case B". In the absence of dust, for reasonable nebular temperatures of $5\times 10^{3}-2\times 10^{4}$ K and electron densities $\rm{n_{e}} <10^{4}  ~\rm{cm^{-3}}$ the expected integrated $\rm{Ly\alpha}/\rm{H\alpha}$ ratio for case B recombination should range between $8.1-11.6$ \citep{Hummer87}. In order to remain consistent with the broader literature, we use the conventional case B ratio of 8.7 set by \citet{Hu1998} (for further discussion, see \citealt{Henry2015, Hayes2015, Trainor2015}).

Note that this ratio assumes spatially integrated measurements with apertures large enough to capture the full Ly$\alpha$ flux as the Ly$\alpha$ can scatter and spatially diffuse while H$\alpha$ cannot. If the spatial aperture does not encompass all the Ly$\alpha$, the measured Ly$\alpha$/H$\alpha$ may be considered to be a lower limit to the true value.

In the second case, the  Ly$\alpha$ emission is produced as mostly neutral hydrogen gas absorbs Ly$\alpha$ and doppler-shifted Balmer continuum photons (``photon-pumping") from the broad line region of QSO A and re-emits them as Ly$\alpha$ photons into our line of sight. In this scenario, we would expect little to no H$\alpha$ to be produced, i.e., we would expect to obtain only a lower limit on the $\rm{Ly\alpha}/\rm{H\alpha}$ ratio. If such a lower limit is at least 12 or so, this scenario could then be distinguished from the recombination case.

\subsubsection{Evidence for a detection of the Slug Nebula's H$\alpha$ emission}
\label{sec:therealHa}
In order to differentiate between these two production mechanisms, it is important to ensure that our H$\alpha$ detection in the MOSFIRE N1 spectrum originated from the Slug Nebula. As mentioned in \S \ref{sec:HaRobust}, despite of the presence of a bright, imperfectly subtracted sky line at the location of our H$\alpha$ detection there are several points that help support the idea that the measured H$\alpha$ flux is indeed emission from the Slug. 

\begin{enumerate}
\item \textbf{There is a significant H$\alpha$ emission}: The H$\alpha$ flux we measured corresponds to a 5.6$\sigma$ detection if we use our empirical sky noise estimator. If we instead use the error file from the MOSFIRE DRP to estimate the noise, the SNR doubles. In either case, the detection is significant despite being on top of a sky line. Though the H$\alpha$ flux we measure is not clearly visible in the unsmoothed 2D MOSFIRE spectrum shown in Figure \ref{fig:Mosfire2Dspec}, once the spectrum is smoothed, as shown in Figure \ref{fig:Mosfire2Dspec_smoothed}, the emission line becomes evident.

\item \textbf{The H$\alpha$ emission is located where it is expected}: When determining the aperture in which to measure the H$\alpha$ flux, we determined the spatial centroid and width solely from the NB image. It is therefore notable that the only significant emission besides QSO A and compact source ``D" in the region is located within these independently derived spatial constraints. In addition, it is striking that despite our priorless search for the velocity centroid of the H$\alpha$ emission, it coincides so well with the velocity centroid of the Ly$\alpha$ emission that we derived from the LRIS slit. 

Note that while the LRIS slit does not match the orientation of the MOSFIRE N1 slit, they do have an overlap region at point ``X", where the velocity centroid of the Ly$\alpha$ and H$\alpha$ are very closely matched. In addition, the LRIS kinematics shown in Figure \ref{fig:LyaKinematicsALL} indicate a constant velocity centroid within each region. Since the MOSFIRE N1 slit goes squarely through region 2 of the Nebula, it is unlikely that the Ly$\alpha$ velocity centroid within a MOSFIRE N1-like slit would differ much from our observed LRIS kinematics. 
  
\item \textbf{The H$\alpha$ emission looks like what we would expect from the Ly$\alpha$}: In order to verify that our observed H$\alpha$ emission looked reasonably similar to what we would expect from the Slug Nebula, we visually compared it to two simple emission prediction models based on the LRIS Ly$\alpha$ emission and an assumption of case B recombination radiation. 

As described in \S \ref{sec:HaRobust} and shown in Figure \ref{fig:FakeSources}, we found that our observed H$\alpha$ detection is visually consistent with a compact 2D gaussian emission model ($\sigma_{\rm{vel}}$=181 km/s and $\sigma_{\rm{spat}}$=18 kpc) with a total flux that is the same as our observed H$\alpha$ flux.  

\end{enumerate}

Thus, there is significant H$\alpha$ emission at a location consistent with that of the Ly$\alpha$ emission of the Slug Nebula and that it looks similar to what would be expected assuming a relatively narrow H$\alpha$ velocity distribution produced by recombination emission. These facts together indicate that our MOSFIRE H$\alpha$ flux is very likely a true detection of the Slug Nebula H$\alpha$ emission. However, the only way to definitively confirm the H$\alpha$ detection at much 
higher significance level would require observations that are not affected by sky-lines, i.e. from space using the James Webb Space Telescope.

\subsubsection{The Fluorescent Nature of the Slug Nebula's Emission}
\label{sec:fluorescentnature}

The ratio of the Ly$\alpha$ flux, measured in \S \ref{sec:Lyaflux}, to the corresponding H$\alpha$ flux, calculated in \S \ref{sec:Halphaflux}, allows us to determine which mechanism is primarily responsible for powering the Slug Nebula emission. We find a ratio of: 
\begin{equation}
\frac{\rm{F_{Ly\alpha}}}{\rm{F_{H\alpha}}} = 5.5\pm1.1
\label{eqn:ratio}
\end{equation}

Despite the large uncertainties and the limitations of our current observations, our measured value of $\rm{F_{Ly\alpha}}$/$\rm{F_{H\alpha}}$ is clearly much lower than the expected ratio of $\rm{F_{Ly\alpha}}$/$\rm{F_{H\alpha}}>12$ if the Ly$\alpha$ emission of the Slug Nebula were primarily being produced via ``photon-pumping" or scattering of the quasar broad line region. Rather, it is remarkably close to the ``standard" case B recombination ratio of 8.7. If, as discussed in \S \ref{sec:Haaperture} and \S \ref{sec:Halphaflux}, the H$\alpha$ flux is biased slightly high due to our aperture selection, the flux ratio could be as large as $\rm{F_{Ly\alpha}}$/$\rm{F_{H\alpha}}$= 6.9$\pm 1.1$, driving it even closer to the canonical 8.7 value. Observing Ly$\alpha$ to H$\alpha$ emission ratios that are so close to those expected for case B recombination, implies that the gas in the Slug Nebula must be mostly ionized, presumably by QSO A, optically thick to Ly$\alpha$ photons, and producing the fluorescent Ly$\alpha$ and corresponding H$\alpha$ emission in-situ as the gas recombines. Of course, some small contribution due to ``photon-pumping" or scattering from the quasar broad line region cannot be excluded. 

In studies of Ly$\alpha$ emitting galaxies, it is customary to interpret this ratio in terms of the Ly$\alpha$ escape fraction ($f_{esc}$). The Ly$\alpha$ escape fraction compares the ratio of observed $\rm{F_{Ly\alpha}}/\rm{F_{H\alpha}}$ (where F$_{H\alpha}$ is generally dust corrected) to the ideal case B recombination value of 8.7 (see i.e. equation 2 of \citealt{Atek2009}). If we convert our measurement of the $\rm{F_{Ly\alpha}}/\rm{F_{H\alpha}}$ ratio from equation \ref{eqn:ratio} into a Ly$\alpha$ escape fraction, it would correspond to $f_{esc} \sim 63\%$. This value is in keeping with the escape fractions found for Ly$\alpha$ selected galaxies at redshifts of z$\sim2-3$, which range from a few percent to over 100\%, but are typically $\sim 30$\%   (e.g. \citealt{Hayes2010, Steidel2011, Erb2014, Trainor2015, Matthee2016}). The presence of dust is often used to explain escape fractions that are below 100\%, since dust preferentially destroys Ly$\alpha$ as compared to H$\alpha$.\citet{Hayes2010, Steidel2011}, and to a lesser extent \citet{Matthee2016}, all observe that $f_{esc}$ is anti-correlated with dust attenuation.

However, it is very important to remember that the Slug Nebula is not a Ly$\alpha$ galaxy. Rather, it is a very massive reservoir of cool gas that spans over 450 kpc, has no detected stellar continuum component, and as discussed in \S \ref{sec:DiscussLyaKinematics}, has kinematics that are inconsistent with being a massive rotating disk. Therefore, as discussed in \citet{Cantalupo2014}, the Slug Nebula is likely a filamentary structure in the IGM, and we do not expect significant amounts of dust to be present on these intergalactic scales. Indeed, the non-detection of metal emission, from \ion{C}{IV}[1549] \citep{FAB2015} suggests that the metallicity of the Slug is not as high as in the ISM of high-redshift galaxies.

Another explanation for the $f_{esc}<100\%$ observed in Ly$\alpha$ emitting galaxies was proposed by \citet{Steidel2011}. As they point out, it is not necessary to destroy Ly$\alpha$ photons to affect the Ly$\alpha$ flux measurement. Notably, resonant scattering causes the Ly$\alpha$ photons to diffuse spatially outwards while leaving the non-resonant H$\alpha$ unaffected. Therefore, an aperture that encompasses all of the H$\alpha$ emission will likely be missing a significant amount of the Ly$\alpha$, leading to measured escape fractions that are less than 100\%. 

This scattering of Ly$\alpha$ photons to larger spatial scales is probably the dominant effect contributing to why our measured Ly$\alpha$ flux is below what we would expect for case B recombination. Since we are measuring the Ly$\alpha$ flux corresponding to the Night 1 slit by integrating the Ly$\alpha$ flux within a pseudo-slit region of the NB image, we are likely missing a significant fraction of the Ly$\alpha$ photons produced in this bright region, particularly those that are scattered by more than the 1$\arcsec$ slit width. This explanation is further supported by the fact that we see possible radiative transfer effects playing a role in producing a Ly$\alpha$ spectral width that is broadened compared to that of the H$\alpha$. In this case, we would expect higher Ly$\alpha$ to H$\alpha$ ratios in the outer, fainter regions 
of the Slug Nebula that are currently not covered by our spectroscopic slit, which was centered on the brightest emission. Deep H$\alpha$ narrow-band or integral field spectroscopic observations would be needed to confirm this scenario.

As discussed in \citet{Borisova2016}, another conceivable contribution to the lower than expected Ly$\alpha$ flux could be ``filter-loss" effects. These filter losses occur when a portion of the broad Ly$\alpha$ emission falls outside of the peak transmission of the NB filter. However, if we compare the measured transmission curve of the NB filter in the laboratory to the Ly$\alpha$ kinematics from our LRIS spectrum, assuming that these kinematics are similar to those that would have been observed using the N2 slit, we find that the Ly$\alpha$ emission coincides well with the NB filter peak transmission and that any filter-losses would be too small to explain the lower than expected Ly$\alpha$ to H$\alpha$ ratio. 

The recombination nature of the Slug Nebula's emission has important implications for the conditions of the gas on intergalactic and circumgalactic scales around quasars. As discussed in detail in \citet{Cantalupo2014}  and \citet{FAB2015} (also see \citealt{Cantalupo2017} for a review) the large Ly$\alpha$ (and H$\alpha$) SB of the Slug in the recombination case would imply very high gas densities ($n>1$ cm$^{-3}$) that can only be explained by large clumping factors (C$\sim$1000, and therefore small volume filling factors) given the large intergalactic scales associated with the emission. In addition, the indications that Ly$\alpha$ is being radiatively broadened due to radiative transfer effects, suggest that the gas is highly ionised but not completely optically thin to Ly$\alpha$ radiation produced by recombination. This would imply a neutral hydrogen column density significantly above 10$^{14}$ cm$^{-2}$ and will help future studies to further constrain the ionisation parameter,
total column densities, and volume densities of the gas.

\subsection{Elucidating the Nature of Compact Sources ``C" and ``D" }
\label{sec:DiscussCompact}
In \S \ref{sec:compactsources}, we calculated the H$\alpha$ and \ion{N}{II}[6583] fluxes for compact sources ``C" and ``D", which we can use to surmise the origin of the H$\alpha$ emission. Cantalupo et al., (in prep.) modeled the UV continuum emission of compact source ``C" using Starburst99 \citep{Starburst99} and found that the galaxy was consistent with have little to no dust and a star formation rate (SFR) of $\approx2-3~ \rm{M_{\odot}}$/yr. We can convert this star formation rate into a predicted H$\alpha$ flux by using the classic conversion of SFR to H$\alpha$ luminosity from \cite{Kennicutt98}. In this way, we calculate that a SFR$=3\rm{M_{\odot}}$/yr corresponds to an expected H$\alpha$ flux of $\rm{F_{H\alpha,expected}}=9.0\times10^{-18}$ erg/cm$^2$/s. Comparing this expected flux to the observed H$\alpha$ flux measured in \S \ref{sec:compactsources}, we find that the observed flux is 4.8 times higher than what would be predicted from star formation alone when using the \citet{Kennicutt98} relation. 

We can perform a similar analysis on compact source ``D". Since we do not have a UV spectrum of compact source ``D", we cannot do the full modeling of its UV continuum emission, as we did for compact source ``C". Instead, we can attempt to rescale the SFR we computed for compact source ``C" to one for compact source ``D" by comparing their UV continuum fluxes as determined from the V-band photometry (see Figure \ref{fig:SlitPA}). We find that $\rm{F_{UV,C}}/\rm{F_{UV,D}}\approx 2.9$, suggesting a SFR for compact source ``D" of $\sim 1\rm{M_{\odot}}$/yr. It is important to note, however, that this method implicitly assumes that compact source ``D", like ``C", has little to no dust extinction in the UV continuum. A SFR of $1\rm{M_{\odot}}$/yr corresponds to a predicted H$\alpha$ flux of $\rm{F_{H\alpha,expected}}=3.0\times10^{-18}$ erg/cm$^2$/s for compact source ``D", which as in the case of compact source ``C" is a factor of $\rm{F_{H\alpha, observed}/F_{H\alpha, expected}} = 22$ times lower than the observed H$\alpha$ flux.

Both of these comparisons rely on the classic \citet{Kennicutt98} value based on the typical conditions of star-formation in nearby, massive galaxies. However, interpreting the additional flux H$\alpha$ as evidence of a contribution of ionizing radiation from a source other than star-formation is not a secure conclusion for many low-mass and vigorously star-forming galaxies. A classic method of interpreting the ionization state of a galaxy is the N2-BPT diagram \citep{Baldwin81,Veilleux1987} which compares the ratio of \ion{N}{II}/H$\alpha$ to \ion{O}{III}/H$\beta$. When examining this relation for low-redshift systems, \citet{Brinchmann2008b, Brinchmann2008a} find a tail of star-forming systems with small \ion{N}{II}/H$\alpha$ and high \ion{O}{III}/H$\beta$, generally with high specific star-formation rates ($10^{7} - 10^{8}$ years$^{-1}$.)  Such galaxies are common in surveys of star-forming systems at $z\sim2$, yielding a N2-BPT diagram populated with extreme ratios of  \ion{N}{II}/H$\alpha$ \citep{Nakajima2013,Maseda2013,Steidel2014,Shapley2015,Holden2016,Trainor2016,Strom2017}. To produce these extreme ratios requires a much harder ionizing flux than typically produced star-forming regions in, for example, the Milky Way. One method of producing these would be the nearby QSO, but the frequency of these galaxies outside of the neighborhoods of QSO points to different conditions of star-formation such as is discussed in, for example, \citet{Kewley2013}, \citet{Steidel2016}, and \citet{Eldridge2017}.

\section{Conclusions}
\label{sec:Conclusions}
The recent discovery of Enormous Ly$\alpha$ Nebulae (ELAN, also referred to as Giant Ly$\alpha$ Nebulae) around quasars has opened up a new observational window to study intergalactic gas in emission on scales of several hundred kpc  around massive galaxies at high redshift (see e.g., \citealt{Cantalupo2017} for a review). 
The Slug Nebula is one of the largest and most luminous among the ELAN discovered to date,
extending over 450 physical kpc around the bright quasar UM287 at z$=$2.283 \citep{Cantalupo2014}
with a very high Ly$\alpha$ surface brightness. 
Depending on the Ly$\alpha$ emission mechanism, these high SB values would imply either 
a ``clumpy" and mostly ionised medium (in the case of recombination radiation) or large column densities
of neutral gas (in the case of ``photon-pumping" or scattering radiation from the quasar broad line region emission),
as discussed in \citet{Cantalupo2014}. 

In order to clearly distinguish between these two scenarios,  we searched for the non-resonant 
hydrogen H$\alpha$ emission from the brightest part of the Slug by means of deep Keck/MOSFIRE long-slit
spectroscopic observations. In addition, we obtained a deep, moderately high-resolution Ly$\alpha$
 Keck/LRIS spectrum in order to guide our H$\alpha$ emission search in the spectral direction 
 and to study the detailed kinematics of the Nebula.

\begin{enumerate}

\item Compared to previous lower-resolution and lower signal-to-noise Ly$\alpha$ spectral studies, our LRIS
observation of Ly$\alpha$ emission revealed a more complex kinematic pattern than a simple, giant rotating disk
\citep{Martin2015}. Instead, as presented in \S \ref{sec:LRISkinematics} and discussed in \S \ref{sec:DiscussLyaKinematics}, these kinematics seem more consistent with the presence of at least two structures that are clearly separated in velocity space.

\item We then independently analyzed the H$\alpha$ spectrum obtained by Keck/MOSFIRE. By optimizing the spectral aperture size and velocity centroid using a curve-of-growth approach, we found an H$\alpha$ detection of F$_{\rm{H\alpha}}= 2.62\pm 0.47 \times 10^{-17}$ erg/cm$^2$/s with a significance of $\sim5.6\sigma$,  at a velocity of -511 km/s from the systemic redshift of the quasar UM287 (z=2.283) (see \S \ref{sec:Haemission} for more details). Such a detection is exactly at the expected velocity and spatial location obtained from the Ly$\alpha$ LRIS spectrum and NB image, respectively, reinforcing the reliability of the detected emission.

\item The observed H$\alpha$ signal overlaps with residuals from a relatively bright IR sky line, reducing the overall signal-to-noise ratio and hampering the possibility of a detailed kinematic analysis of this emission. However, our curve-of-growth analysis in \S \ref{sec:Haaperture} suggests that the H$\alpha$ emission could be significantly more narrow (181 km/s) than its Ly$\alpha$ counterpart (418 km/s). This possible broadening of the Ly$\alpha$ emission as compared to the H$\alpha$ emission would naturally be produced by resonant scattering of Ly$\alpha$ photons if the Nebula were optically thick to the Ly$\alpha$ radiation, thus implying N$_{\mathrm{HI}}>10^{14}$ cm$^{-2}$.

\item The most important result from our observations is the direct measurement of the Ly$\alpha$ to H$\alpha$ ratio in the region covered by our MOSFIRE N1 slit. We found the ratio $\rm{F_{Ly\alpha}}/\rm{F_{H\alpha}}$ to be 5.5$\pm1.1$ $+1.4$ (sys), see \S \ref{sec:Haaperture} for a discussion of the systematic error. Since ``photon-pumping" or scattering emission from the quasar broad line region contribute Ly$\alpha$ photons without producing any corresponding H$\alpha$ photons, we would expect these emission mechanisms to result in very high values of $\rm{F_{Ly\alpha}}/\rm{F_{H\alpha}}$ that would be well above the expected case B recombination (8.7 for total integrated emission or slightly lower for a slit observation like our own). 

Therefore, the fact that the observed Ly$\alpha$ to H$\alpha$ is this close to the expected case B recombination value suggests that any contribution to the Ly$\alpha$ emission from these alternate emission mechanisms should be negligible and that the dominant source of Ly$\alpha$ emission for the Slug Nebula is recombination radiation. As derived in \citet{Cantalupo2014}, \ion{H}{I} column densities above N$_{\mathrm{HI}}\sim10^{19}$ cm$^{-2}$ are expected to have a significant Ly$\alpha$ flux contribution due to ``photon-pumping" or scattering from the quasar broad line region. Thus, our Ly$\alpha$ to H$\alpha$ flux ratio places an upper limit on the \ion{H}{I} column density of N$_{\mathrm{HI}}<10^{19}$ cm$^{-2}$.   
\end{enumerate}

Taken as a whole, the above conclusions imply that the emission from the Slug Nebula is powered by recombination with minimal contributions due to
scattering of ex-situ Ly$\alpha$ photons. Thus, the intergalactic and circumgalactic medium around UM287 must be highly ionized, with an \ion{H}{I} column density between $10^{14}$ cm$^{-2}$ to  $10^{19}$ cm$^{-2}$. Considering the work of \citet{Cantalupo2014} and \citet{FAB2015}, this suggests that the Slug Nebula emission requires the presence of high density gas structures (``clumps") with a small volume filling factor.  Though the exact gas density distribution is not well constrained, these ``clumps" could be the high density tail of a very broad gas distribution (Cantalupo et al, in prep.). 

Despite the technical challenges and limitations of extended, faint emission spectroscopy in the IR, our result demonstrate the potential of H$\alpha$ intergalactic fluorescent observations at high-redshift. Future surveys from space-based observatories such as JWST that do not suffer from the presence of sky-lines would be necessary for a significant step forward for the H$\alpha$ study of the Slug Nebula and for other enormous Ly$\alpha$ nebulae at high redshift. 

\section*{Acknowledgements}
The authors would like to thank the referee for their incredibly helpful and thoughtful review of this paper. The National Science Foundation (NSF) grants AST-1010004 and AST-1412981 helped support C.N.L., S.C. and Jason Prochaska, who participated in the data collection, reduction and analysis. S.C. gratefully acknowledges support from Swiss National Science Foundation grant PP00P2\_163824. Support for this work was also provided by NASA through grant HST-AR-13904.001-A (P.M.) and the National Science Foundation Graduate Research Fellowship under Grant No. NSF DGE1339067 (C.N.L.). P.M. also acknowledges a NASA contract supporting the WFIRST-EXPO Science Investigation Team (15-WFIRST15-0004), administered by GSFC, and thanks the Pr\'{e}fecture of the Ile-de-France Region for the award of a Blaise Pascal International Research Chair, managed by the Fondation de l'Ecole Normale Sup\'{e}rieure. The data presented herein were obtained at the W. M. Keck Observatory, which is operated as a scientific partnership among the California Institute of Technology, the University of California and the National Aeronautics and Space Administration. The Observatory was made possible by the generous financial support of the W. M. Keck Foundation. The authors wish to recognize and acknowledge the very significant cultural role and reverence that the summit of Maunakea has always had within the indigenous Hawaiian community.  We are most fortunate to have the opportunity to conduct observations from this mountain.

\bibliographystyle{mnras}

\label{lastpage}
\end{document}